\documentclass[prb,nobibnotes,aps,showkeys]{revtex4} 

\usepackage{times,helvet}

\usepackage{graphicx}
\usepackage{subfigure}
\usepackage{afterpage}

\usepackage{amsmath}
\usepackage{amssymb}
\usepackage{dsfont}
\usepackage{mathrsfs}

\usepackage{fancyhdr}

\makeatletter
\renewcommand\@dotsep{2}
\makeatother

\begin{document}

\title{The interplay between discrete noise and nonlinear chemical kinetics in
  a signal amplification cascade}

\author{Yueheng Lan}
\author{Garegin A. Papoian}
\email[Electronic mail: ]{gpapoian@unc.edu}
\affiliation{Department of Chemistry, University of North Carolina at Chapel Hill, NC 27599-3290}

\date{\today}

\begin{abstract}

  We used various analytical and numerical techniques to elucidate signal
  propagation in a small enzymatic cascade which is subjected to external and
  internal noise. The nonlinear character of catalytic reactions, which
  underlie protein signal transduction cascades, renders stochastic signaling
  dynamics in cytosol biochemical networks distinct from the usual description
  of stochastic dynamics in gene regulatory networks. For a simple 2-step
  enzymatic cascade which underlies many important protein signaling pathways,
  we demonstrated that the commonly used techniques such as the linear noise
  approximation and the Langevin equation become inadequate when the number of
  proteins becomes too low. Consequently, we developed a new analytical
  approximation, based on mixing the generating function and distribution
  function approaches, to the solution of the master equation that describes
  nonlinear chemical signaling kinetics for this important class of
  biochemical reactions. Our techniques work in a much wider range of protein
  number fluctuations than the methods used previously. We found that under
  certain conditions the burst-phase noise may be injected into the downstream
  signaling network dynamics, resulting possibly in unusually large
  macroscopic fluctuations. In addition to computing first and second moments,
  which is the goal of commonly used analytical techniques, our new approach
  provides the full time-dependent probability distributions of the colored
  non-Gaussian processes in a nonlinear signal transduction cascade.
\end{abstract}

\maketitle

\section{INTRODUCTION}

Biochemical signaling networks mediate information flow in cells, regulating
important cellular processes such as cell metabolism, motility, gene
expression and cell cycle\cite{st02gomp,den95p}. A quantitative understanding
of signal transduction is essential for realizing the larger goal of modeling
the cell. Until recently, biochemical reaction networks were analyzed mainly
with the help of chemical kinetics equations, exploring such issues as
robustness, sensitivity, and
bistability\cite{math02hei,opt04cha,han01ex,nick04ph,h03ecoli}. This
deterministic description, though valid in the asymptotic limit of a
macroscopic system (as for a reaction in organic chemist's test-tube), fails
when the number of reacting proteins is too small. This is often the case in
the cell, which is a mesoscopic system. Therefore, instead of a smooth
deterministic course, the fundamentally random nature of chemical reactions
results in ``noisy'' reaction trajectories in individual cells. The resulting
heterogeneous response of an ensemble of cells to a particular external
signal\cite{wein05st} necessitates going beyond chemical kinetics, relying
instead on the theory of stochastic
processes\cite{cao04eff,tan04mod,tur04st,dm05del} to describe signaling
dynamics.

Previous theoretical and experimental results have strongly suggested that
stochasticity is an important component in the dynamical processes in a
cell\cite{muk04st,jef02tr,tan04mod,sasai03stch,muk02att,rao02n,hol05st,coh05fl}.
For instance, in signal transduction, fluctuations induce not only
quantitative changes of threshold values\cite{walcz04stgn} but also
qualitative changes such as oscillations, transitions between different stable
states, and stochastic resonances that may increase sensitivity or stability
of the cellular signal
processing\cite{paul00st,paul00rd,sh05noise,bark99cir,kurt95st,jos01n}. The
resulting large variability in cell-to-cell response to the same external
stimulus\cite{wein05st,kor04var} allows a cell colony to adapt to varying
environment.

The connection between deterministic and stochastic chemical kinetics is
analogous to that between classical and quantum
mechanics\cite{wang94sur,wang96ins}. In particular, the number of degrees of
freedom in stochastic description of chemical kinetics is immensely larger
compared with the deterministic description. Consequently, the equations of
stochastic chemical kinetics are difficult to solve, and mainly numerical
results have been discussed in prior work. Small sizes and intrinsic
complexities of cells allow for the emergence of large fluctuations that are
coupled with other rich dynamical processes. Thus, the resulting formidable
complexity of these dynamical systems has hindered obtaining a complete
solution to the problem of stochastic dynamics in biochemical reaction
networks. How to construct approximate analytical solutions to understand
qualitatively the complex signaling dynamics is a challenging problem which
stands in the research frontier of non-equilibrium statistical mechanics and
dynamical systems\cite{sm05inf,walcz04stgn}.

A number of tools have been developed to deal with the randomness in chemical
reactions\cite{van92st,gar02han,ris84fok,sasai03stch,wang02}. The Gillespie
simulation algorithm\cite{gill77ext,gil01app,jer05sim}, the Langevin equation
and the Fokker-Planck equation
\cite{van92st,lin04hay,tao05st,jer05sim,elf03fast,gil00lan} are among the
most commonly used. These methods have been applied fruitfully to study signal
transduction
processes\cite{cao04eff,tan04mod,tur04st,gen04ch,shnerb01aut}. However, a
number of constraints limit the applicability of these methods. Gillespie
simulation is essentially a Monte Carlo algorithm which simulates chemical
reactions as a series of transitions between reaction states with transition
rates determined from microscopic kinetics. One simulation corresponds to one
possible reaction trajectory. When the system is large, one or two reaction
paths provide qualitative or even quantitative information of the system
dynamics. However, to get good statistics, a large number of paths are
required when the system size is moderate or small. This, in many cases, may
be computationally expensive. Moreover, although the computational cost is
dominated by the fastest reaction in a cascade, the time scale of interest is
likely given by the slowest reaction, perhaps many orders of magnitude slower
than the fast reaction. In this situation, commonly encountered in biological
signaling networks, producing a single trajectory is computationally
forbidding. Under special conditions these simulations can be
accelerated~\cite{gil03st,gil01app,gil00lan,cao04eff,jac04b,gil03imp,eric02app,vas04ad,puc04bdg},
but for a general reaction network this is still an active area of
research. On the other hand, a qualitative understanding of the reaction
network dynamics is essential for learning control of biochemical processes in
the cell\cite{keen05len,h03ecoli}. It is difficult to extract such
understanding from Gillespie simulations alone, without the guidance from a
complementary analytical model.

A number of approximate analytical techniques are used to solve the chemical
master equation. Among the most commonly used ones, the linear noise
approximation\cite{tao05st,lin04hay}, is an effective weak-noise expansion
based on the fact that fluctuations are of order $\sqrt{\Omega}$ for a system
of size $\Omega$. It works well for large systems where fluctuations are small
such that the probability distribution is centered narrowly around
deterministic orbits determined from chemical kinetics\cite{van92st}. However,
molecular discreteness and large fluctuations in cellular biochemical
reactions, combined with nonlinear effects, may generate strong correlations
along a pathway, leading to the formation of characteristic patterns both in
time and
space\cite{hol05st,coh05fl,ku01sel,keen01df,lem05st,kul04pat,h03ecoli}.

To account for such patterns, we developed an approach which incorporates
large fluctuations, going beyond the commonly-used small noise and continuous
approximations. We focus our efforts on a specific 2-step signaling cascade,
consisting of a unary reaction of receptor activation upstream and a
subsequent binary reaction of enzyme activation downstream. Biochemical
signaling networks are comprised of only few signaling elements, and the
2-step signaling cascade considered in this work is among most fundamental
building blocks. The nonlinearity of the catalytic reaction in the second step
is the main source of difficulties in obtaining a comprehensive analytical
solution to stochastic signal dynamics in this amplification cascade. In this
work, we developed high-quality approximate solutions to the stochastic signal
propagation dynamics in a 2-step cascade, uncovering that fluctuations may
broaden the average chemical kinetics signals such that transient, highly
non-Gaussian distributions may emerge, due to interplay between discrete noise
and nonlinearity. We could not reproduce this effect using the chemical
Langevin equation, since the latter is based on the continuous approximation,
ignoring molecular discreteness.

The paper is organized as follows. After commenting on the strengths and
weaknesses of the traditional modeling techniques, we solve the master
equation for the 2-step cascade with the help of generating functions. Then
the new formalism is used to elucidate how noisy signal may be controlled in
an unbranched signaling pathway. In particular, the upstream fluctuations may
propagate downstream without dissipation, resulting in large downstream
fluctuations even in the limit of a macroscopic size downstream signaling
network. Although we focused in this work on an important, yet specific
enzymatic cascade, the hybrid generating function -- distribution function
technique introduced in this work (Section \ref{sect:smooth}) may be
generalized to treat more complex biochemical pathways. To demonstrate the
usefulness of this hybrid ``smooth distribution'' method, we consider in
Section \ref{sect:smooth}b a self-dimerization of the receptors in the first
step of the cascade activation, which enhances the nonlinear character of the
2-step cascade. In a forthcoming publication we will illustrate the use of
time-dependent basis functions, developed in this work, in the variational
solution of stochastic chemical kinetics equations in signaling cascades
comprised of several steps and containing feedback loops.
 
\section{MASTER EQUATION TREATMENT OF REACTION TRAJECTORY REALIZATIONS}

Signal transduction often starts at the cell membrane, where external ligands,
such as hormones or small peptides, bind and activate cell surface
receptors. In turn, the activated receptors send the signal downstream,
usually by phosphorylating specific proteins inside the
cell\cite{st02gomp}. These proteins then activate other cytosol proteins
further downstream. The process goes on so that the signal propagates through
the cell in a predetermined way. However, the signaling dynamics is not
uniform when protein numbers are small, which is often the case in the burst
phase of the cascade activation. Because of the fundamentally random nature of
chemical reaction dynamics\cite{van92st}, each trajectory realization is
different from others, even when the same initial conditions are used. This
behavior is called trajectory variability in the literature\cite{kor04var}. If
the variation is large, then a stochastic description becomes necessary.

\begin{figure}[tbh]
\includegraphics[width=12cm]{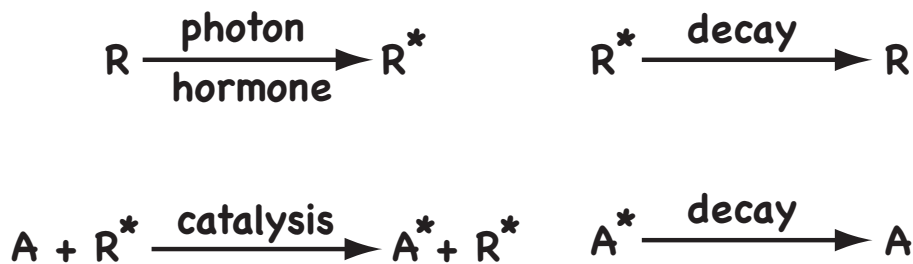}
\caption{An inactive receptor $R$, when activated by a signal, activates
  downstream protein $A$.}
\noindent 
\label{f:react}
\end{figure}

Our long-term goal is to understand qualitatively how large signaling networks
process information. The 2-step amplification cascade, shown in
Fig.~\ref{f:react}, can be regarded as the building block of more complex
cascades. In this simple reaction scheme, without feedback loops, $R$
represents an inactive receptor, which becomes activated into $R^*$ upon
binding of an external ligand (stimulus). When the receptor is activated, it
acts as an enzyme, catalyzing the phosphorylation of the next kinase
downstream ($A+R \to A^*+R^*$) with a rate $\mu$. $A^*$ spontaneously decays
to $A$ with a rate $\lambda$. Although the $R^*$ reaction is unary and
independent of the $A$ reaction, the latter one is binary, making the system
\underline{nonlinear}, thus, different from those usually considered in the
gene regulatory networks\cite{th01in,kie01eff,sw02in,oz02reg}.

\begin{figure}[tbh]
\centering
\includegraphics[height=8cm]{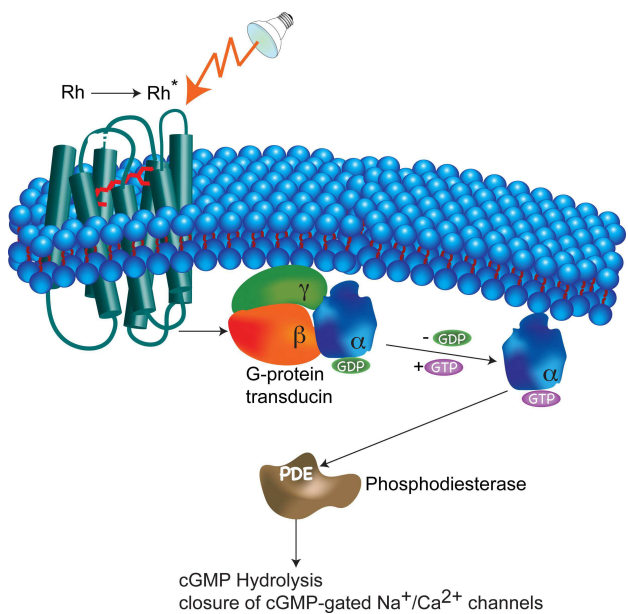}
\caption{A schematic depiction of the visual signal transduction cascade.}
\noindent 
\label{f:phore}
\end{figure}

This simple 2-step cascade is commonly embedded in the onset of a reaction
pathway of many important signaling cascades\cite{pugh92r,sch02com}. For
example, the visual signal transduction pathway, which takes place in retinal
rod cells, is shown in Fig.~\ref{f:phore}\cite{pugh92r,bur01act}. Rhodopsin
receptors (Rh), located in small discs in the outer segments of the rod cell,
contain a light-sensitive molecule, retinal. Upon incidence of photons, the
retinal molecules undergo isomerization, leading to a subsequent activation of
rhodopsin receptors (Rh$^*$). The latter act as a catalyst to replace the GDP
by GTP in a G-protein, called transducin. The next enzyme in the cascade, the
cGMP phosphodiesterase (PDE), is then activated to (PDE$^*$) by the active
G-protein GTP complex. PDE$^*$ hydrolyzes the cGMP, which leads to the closure
of cGMP-gated Na$^+$/Ca$^{2+}$ ion channels and, thus, reduces the influx of
Na$^+$/Ca$^{2+}$ flow. The rod cell becomes hyperpolarized, resulting in less
neurotransmitters being released from the synaptic terminal of the rod
cell. This change is immediately picked up by a second cell and passed as a
signal to the central neural system. In this example, the initiating steps of
the visual signal transduction pathway,
\[
\mbox{Rh} \leftrightarrow \mbox{Rh}^*\,,\quad \mbox{G-GDP}+\mbox{Rh}^* \leftrightarrow \mbox{G-GTP}+\mbox{Rh}^*
\]
are similar to our 2-step cascade. The amazing ability of retina rod cells to
detect single photons, in the presence of external and internal noise, has
been discussed in prior works\cite{sch95pht}. Although understanding visual
signal transduction is not the focus of this work, the formalism of large
fluctuations, developed in the current work, may serve as a basis for further
elucidating this issue.
   
We regard the $R \to R^*$ reaction (Fig.~\ref{f:react}) as a Poisson process
with rate $g$, which corresponds to an abundance of $R$ receptors and scarce
ligand presence. For instance, the light perception in a dark
room\cite{det00en,rk98or} may be described as a Poissonian bombardment of
photons on the retina cell surface. If the intensity of photons is low, there
always exist a sufficient excess of inactivated photon receptors ($R$) such
that the arrival events of photons dominate the process. $R^*$ spontaneously
decays back to $R$ at a rate $k$. The average number of activated receptors
($R^*$) depends both on the incidence rate $g$ and decay rate $k$. If $R \to
R^*$ is considered to be an ordinary first order chemical reaction, instead of
a Poisson process, all methods described below would still apply, with only
minor modifications. Our current investigation is restricted to the $0$-D
treatment of space. This is a good approximation if the reaction network is
confined to a small enough spatial region, having a linear dimension of
$\zeta$ (so-called Kuramoto length\cite{van92st}), such that particles
diffuse across the region fast compared to the typical reaction times. Using
the reaction parameters from published models of the EGF/MAPK signaling
cascade\cite{sch02com,thr04dif,el99prt}, we estimated $\zeta$ to be in the
range from 0.3 $\mu m$ to 5 $\mu m$. In an ongoing work we are incorporating
an explicit treatment of space into our analysis.

\begin{figure}[tbh] 
\centering
\includegraphics[width=8cm]{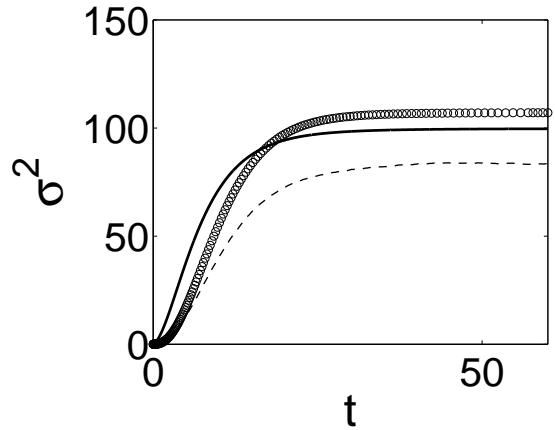}
\caption{Time evolution of the variance of $A^*$ computed from Eq.~\ref{eq:m1}
  ({\it solid line}), Langevin equation ({\it dashed line}) and
  Eq.~\ref{eq:ch2m} ({\it circles}).  For
  $(g,k,\mu,\lambda)=(0.2,0.1,0.02,0.15)$ with initial condition
  $(N_R,N_{R^*},N_A,N_{A^*})=(100,0,100,0)$.  }
\label{f:omgill}
\end{figure}

When the number of protein copies in the signaling network is small, the
evolution of the averages, described by ordinary chemical kinetics, is
inadequate to characterize the system dynamics. An example of the evolution of
the $A^*$ protein number variance, shown in Figure \ref{f:omgill}, indicates
that the amplitude of the fluctuations is about $10$ in the steady state. This
is significant when compared to the deterministic average of about
$20$. Clearly, a stochastic description is required in this case. The chemical
master equation\cite{van92st} is a starting point for studying large
variations of individual stochastic trajectories from the average path. For
the 2-step cascade described above (Fig.~\ref{f:react}), we denote by $P(m,n)$
the probability of having $m$ copies of $R^*$ and $n$ copies of $A$ at a
particular time point. The time evolution of this probability distribution is
determined by the following master equation
\begin{widetext}
\begin{eqnarray}
\frac{dP}{dt}(m,n) &=& \mu[-mnP(m,n)+m(n+1)P(m,n+1)]+\lambda[-(N-n)P(m,n)
 \nonumber \\
&+& (N-n+1)P(m,n-1)]+ g[-P(m,n)+P(m-1,n)]\nonumber \\
&+& k[-mP(m,n)+(m+1)P(m+1,n)]
\,, \label{eq:m1}
\end{eqnarray}    
\end{widetext}
which expresses the transition rates of probabilities from time $t$ to time
$t+dt$ in terms of the probability distribution at time $t$.
The sum of $A$ and $A^*$ is taken to be constant throughout the reaction ($N$
in Eq.~(\ref{eq:m1})). Another way to think about the chemical reaction
dynamics given by Eq.~(\ref{eq:m1}) is to view it as a random walk on a
two-dimensional lattice of integer coordinates $m$ and $n$ with
position-dependent jump probabilities\cite{van92st}.

Since the master equation provides a full description of stochastic chemical
process, solutions of the set of coupled ordinary differential equations in
Eq.~(\ref{eq:m1}) provide all necessary information to analyze the signaling
dynamics. However, an exact analytical solution for the system of ODEs in
Eq.~(\ref{eq:m1}) is not known. Direct numerical integration is also difficult
due to the enormous number of ODEs (10$^4$-10$^8$) for a cascade that may
contain up to 10$^2$-10$^4$ proteins of each species. Simulations based on the
Gillespie algorithm may be used to estimate $P(m,n)$ in Eq.~(\ref{eq:m1}),
however, they are computationally inefficient and are hard to interpret, as
discussed earlier. Thus, to gain qualitative insight into stochastic signal
transduction by the 2-step cascade in Fig.~\ref{f:react}, it is important to
develop an approximate analytical solution to Eq.~(\ref{eq:m1}). 

Physical
considerations are crucially important in making sensible approximations to
the master equation.
Several numerical and analytical
techniques are available to account for the noise effect on signaling dynamics,
including the chemical Langevin 
equation~\cite{gil00lan} and van Kampen's $\Omega$-expansion\cite{van92st}. 
But they are most useful only when the particle numbers are large and
fluctuations are relatively small.
In Appendix A we derive an
$\Omega$-expansion\ for the 2-step cascade (\ref{f:react}). 
In the remaining part of the paper, we develop
alternative solution schemes to solve equations like Eq.~(\ref{eq:m1}), based 
on the generating function approach, to obtain
signaling dynamics in the regime of large fluctuations, where
traditional analytical techniques are no longer applicable. Comparison will be 
made between our method and the results from Langevin equation or $\Omega$-expansion. 

Generating function approaches have been used in prior works to elucidate
stochastic processes in gene regulatory networks\cite{th01in,sw02in}. The
novelty of our work is to extend these techniques to nonlinear chemical
processes in protein signal amplification cascades. The nonlinearity of
chemical reactions greatly enriches the dynamical behavior of signaling
cascades. However, to develop a robust analytical approach to solving
stochastic amplification dynamics, substantial additional difficulties need to
be overcome compared to describing noisy dynamics in linear biochemical
networks.

\section{THE GENERATING FUNCTION APPROACH}
\label{sect:gene}

To treat signal transduction in a wider range of protein numbers in the cell
and to analyze the effects of large fluctuations, we have developed a new
approach, based on generating functions, to solve the master equation. A
generating function encodes probability distributions as its Taylor series
coefficients. As a result, the enormous set of ODEs in Eq.~(\ref{eq:m1}) are
compactified into a single PDE. Thus, the evolution of the probability
distribution can be obtained by solving this one PDE for the generating
function. Since even for medium-sized cascades, there are an astronomical
number of ODEs in the master equation formalism, an approximate generating
function greatly facilitates qualitative and quantitative analysis of
strongly-fluctuating dynamics in biochemical reaction networks. 

As an example,
for the 2-step cascade, we define a generating function through the following
power series
\begin{equation}
\Psi(x,y)=\sum_{m,n} P(m,n)x^m y^n
\,. \label{eq:gendef}
\end{equation}
which satisfies the time evolution equation
\begin{equation}
\frac{\partial \Psi}{\partial t} = (1-y)(\mu x\frac{\partial^2}{\partial
  x \partial y}-\lambda N+\lambda y \frac{\partial}{\partial y})\Psi
  +g(x-1)\Psi-k(x-1)\frac{\partial \Psi}{\partial x}  \label{eq:m4dx}
\,. 
\end{equation}
Our next goal is to develop approximate techniques to solve
Eq.~(\ref{eq:m4dx}). We know that the solution of Eq.~(\ref{eq:m4dx}) is an
analytic function of $x\,,y$ with nonnegative time-dependent coefficients. The
highest derivative in Eq.~(\ref{eq:m4dx}) is $\partial^2 \Psi/\partial
x \partial y$ which reflects the binary chemical reaction between $R^*$ and
$A$. If this term is omitted, Eq.~(\ref{eq:m4dx}) can easily be solved by the
method of characteristics. However, this would completely alter its physical
content. On the other hand, we notice that the generating function of the
$R^*$ distribution $\phi(x)=\Psi(x,1)$ does not depend on the dynamics of $A$
and obeys the following PDE
\begin{equation}
\frac{\partial \phi}{\partial t}=g(x-1)-k(x-1) \frac{\partial \phi}{\partial x}
\,, \label{eq:req}
\end{equation}
which can be solved exactly, resulting in
\begin{equation}
\phi(x)=\exp[\frac{g}{k}(x-1)(1-e^{-kt})]
\,,\label{eq:ysol}
\end{equation}
where the initial condition $N_{R^*}=0$ at $t=0$ was used. The $R^*$
probability distribution, $P(m)$, is given by the coefficients of the series
expansion of Eq.~(\ref{eq:ysol})
\begin{equation}
\phi(x)=\sum_{m=0}^{\infty}\exp(-\frac{g}{k}(1-e^{-kt}))
(1-e^{-kt})^m(\frac{g}{k})^m \frac{x^m}{m!}
\label{eq:ysolexp}
\end{equation}
resulting in, 
\begin{equation}
P(m)=\exp(-(1-e^{-kt})g/k)(1-e^{-kt})^m(\frac{g}{k})^m/m!
\,. \label{eq:rdis}
\end{equation}
Therefore, the time-dependent distribution of $R^*$ is Poissonian and relaxes
to a stationary distribution with the rate $k$. Although this distribution is
generated by both the birth and decay of $R^*$, the relaxation is independent
of birth rate $g$. From Eq.~(\ref{eq:ysolexp}), the average and the variance
of $N_{R^*}$ are also easily calculated, $\langle N_{R^*}\rangle=\langle
\sigma^2 \rangle =(1-e^{-kt})g/k $. 

Next, we build up the cascade by considering also the reactions involving
$A$. In particular, we construct a series expansion\cite{ince} in $x$ with
time-dependent functions $\phi_m(y)$,
\begin{equation}
\Psi(x,y)=\sum_{m=0}^{\infty} \phi_m(y) x^m
\,. \label{eq:newphi}
\end{equation}
Thus, with each state containing $m$ $R^*$'s, we associate a distribution 
of $A$, which may be computed from $\phi_m(y)$.
The new functions $\phi_m(y)$ satisfy
\begin{widetext}
\begin{equation}
\frac{\partial \phi_m}{\partial t}=(1-y)(\mu m \frac{\partial}{\partial y}-\lambda N
+\lambda y\frac{\partial}{\partial y})\phi_m
+km(\phi_{m+1}-\phi_m)+g(\phi_{m-1}-\phi_m)+k\phi_{m+1}
\,.\label{eq:newexp}
\end{equation}
\end{widetext}
Here an infinite hierarchy of coupled linear PDEs are obtained for the unknown
functions $\phi_m(y)$. Eq.~(\ref{eq:newexp}) is exactly equivalent to
the master equation and physical considerations will next be used in finding
good-quality approximate solutions. We start by keeping only the first term in
Eq.~(\ref{eq:newexp}), thus, ignoring the $R \to R^*$ dynamics, and then
incorporate back the omitted terms in an effective way.

\subsection{Time-scale separation modulates noise propagation}

The first term on the right hand side of Eq.~(\ref{eq:newexp}) describes the
birth and death of protein $A$ and the remaining terms describe $R^*$
dynamics.  If the $R^*$ reaction is ignored, the hierarchy of PDEs become
uncoupled. We obtained an exact analytical solution for the resulting PDEs
using the method of characteristics:
\begin{equation}
\phi_{m}^{(0)} (y) = \phi_{m,0}[1+\left(\frac{\lambda}{\lambda+\mu m}
+\frac{\mu m}{\lambda+\mu m}e^{-(\lambda+\mu m)t}\right)(y-1)]^N \,, \label{eq:phi0th1} 
\end{equation}
where the number of $A$'s was taken to be $N$ at $t=0$ and $\phi_{m,0}$ is a
constant, representing the probability of having exactly $m$ $R^*$'s. If, for
example, the number $m$ of $R^*$ is fixed at a particular value $\bar{m}$,
then $\phi_{\bar{m},0}=1\,,\phi_{m}=0$ for $m \neq \bar{m}$. The obtained
generating function indicates that the $A$ distribution is binomial. Note that
the relaxation rate is $\lambda+\mu \bar{m}$, depending on both $\lambda$ and
$\mu$. The solution further simplifies in the limit of long times ($t \to
\infty$).
\begin{equation}
\phi_{\bar{m}}^{(0)} \to  \phi_{\bar{m},0}[1+\frac{\lambda}{\lambda+\mu \bar{m}}(y-1)]^N 
\,, \label{eq:phi0th}
\end{equation}
which is the generating function for the stationary distribution of $A$.

In the real 2-step biochemical cascade, the number of $R^*$'s is, of course,
fluctuating. However, Eq.~(\ref{eq:phi0th1}), with $m$ concentrated at
$\bar{m}$, still constitutes a good approximation when \underline{either} of
the following conditions is satisfied: (i) $R^*$ is characterized by a sharp
distribution centered at $\bar{m}$, which is often the case when the number of
$R^*$'s is large; (ii) the reaction rates for $R^*$ birth and death are much
larger than those for $A$. For a cascade that satisfies condition (i), the
linear noise approximation might be applicable. However, our solution, based
on the generating function approach, provides the full probability
distribution in an analytical form. When condition (ii) is satisfied, the $A
\to A*$ reaction only ``sees'' an average number ($\bar{m}$) of $R^*$. In this
case, our solution is simpler and, perhaps more convenient to use, than the
$\Omega-$expansion solution. Analysis of the cascade dynamics for case (ii)
using Eq.~(\ref{eq:phi0th1}) suggested a possible mechanism for noise
filtering. Even in the case of broad or irregular distribution for $R^*$, if
the fluctuations around the average are fast, the distribution of $A$ is still
well approximated by Eq.~(\ref{eq:phi0th1}), being well-peaked at the average
for large $N$.

\begin{figure*}[tbh] 
\centering
\includegraphics[width=14.cm]{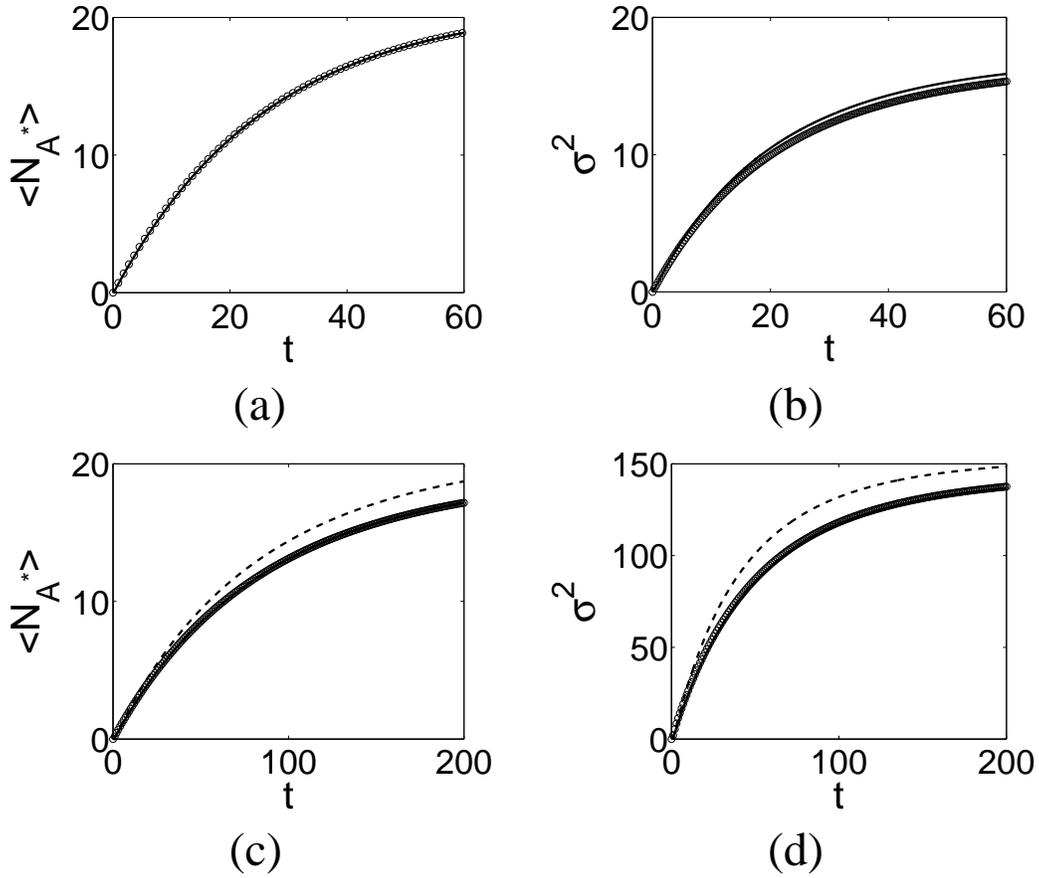}
\caption{Time evolution of the averages ((a) and (c)) and variances ((b) and
  (d)) for $A^*$ obtained from three different calculations: the approximate
  solution Eq.~(\ref{eq:phi0th1}) ({\it circles}), the exact solution ({\it
    solid line} computed from Eq.~(\ref{eq:m1})) and the $\Omega$-expansion
  Eq.~(\ref{eq:ch2m}) ({\it dashed line} probably overlapped by the {\it solid
    line}). The initial condition is
  $(N_R,N_{R^*},N_A,N_{A^*})=(100,0,100,0)$. (a) and (b) have parameter values
  $(g,k,\mu,\lambda)=(20,10,0.004,0.03)$; (c) and (d) have parameter values
  $(g,k,\mu,\lambda)=(0.02,0.01,0.2,1.5)$.  }
\label{f:cas2m12}
\end{figure*}

\begin{figure}[tbh] 
\centering
\subfigure[]{
\includegraphics[width=8.cm]{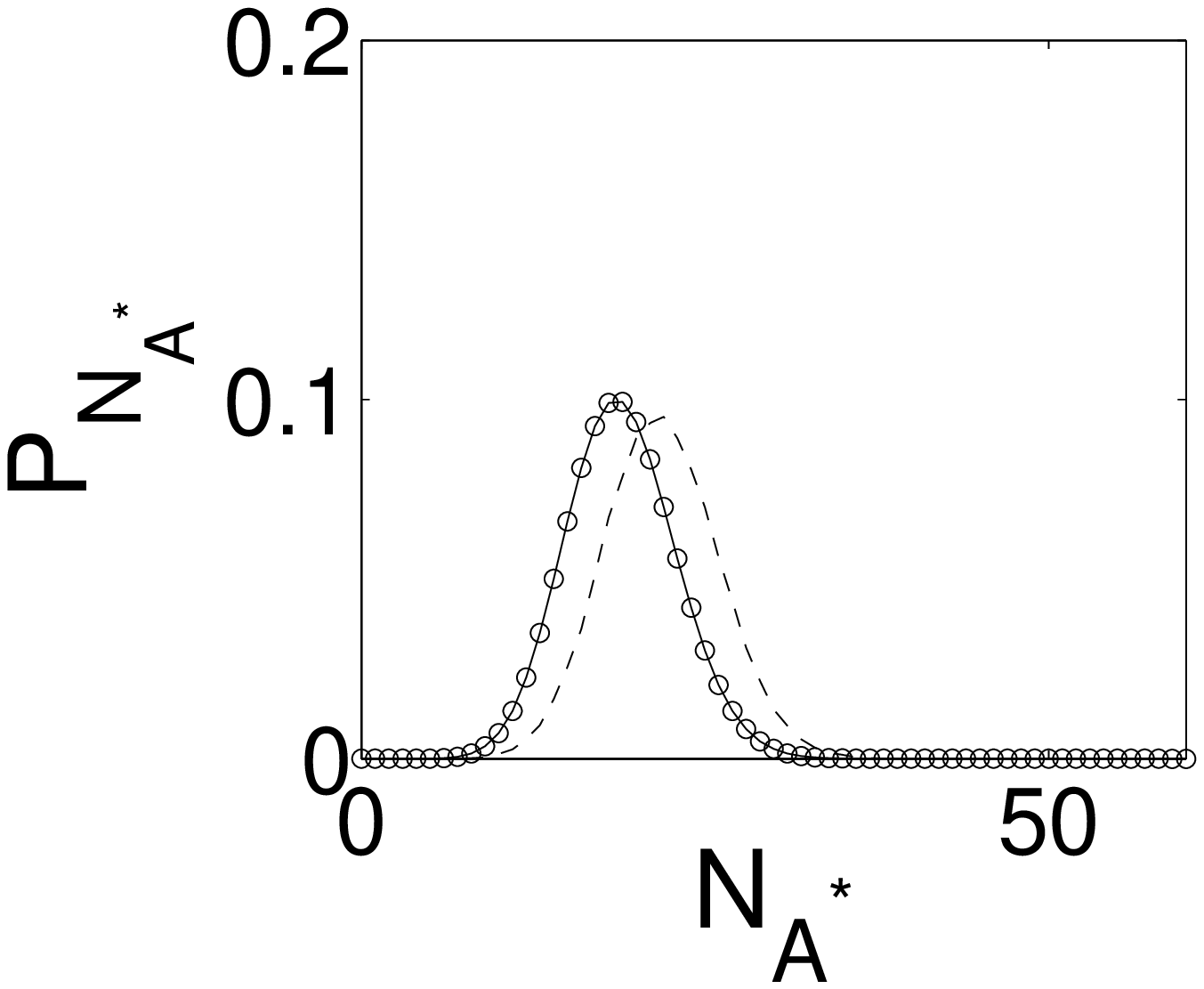}}
\subfigure[]{
\includegraphics[width=8.cm]{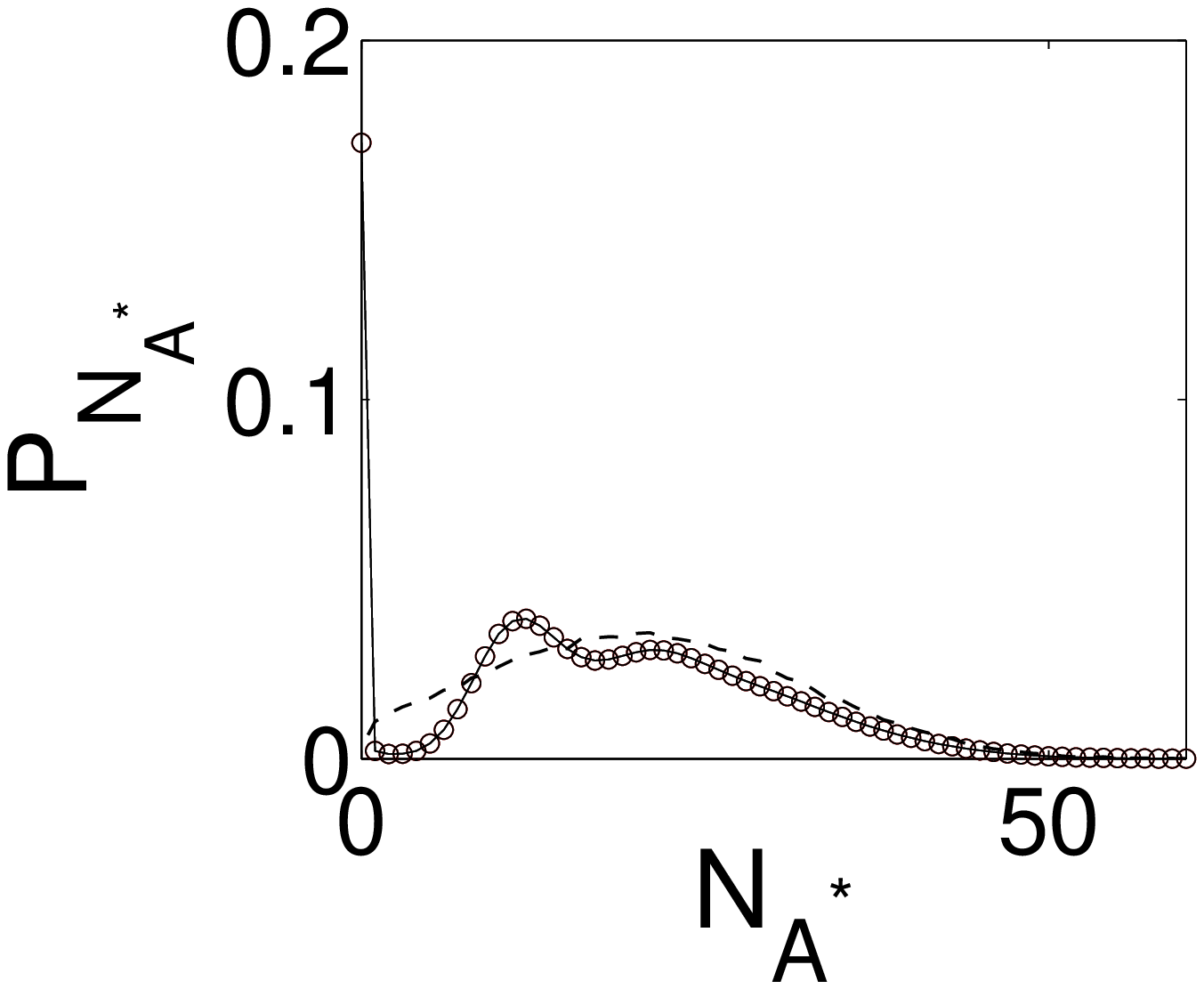}}
\caption{Probability distributions of $A^*$ computed from the approximate
  solution Eq.~\ref{eq:phi0th1} ({\it circles}), Langevin equation
  ({\it dashed line}) and the exact solution
  Eq.~\ref{eq:m1} ({\it solid line}) with initial condition
  $(N_R,N_{R^*},N_A,N_{A^*})=(100,0,100,0)$. (a) The distribution $P(N_{A^*})$
  at $t=60$ with parameter values $(g,k,\mu,\lambda)=(20,10,0.004,0.03)$. (b)
  The distribution $P(N_{A^*})$ at $t=200$ with parameter values
  $(g,k,\mu,\lambda)=(0.02,0.01,0.2,1.5)$.  }
\label{f:cas2d}
\end{figure}

As an example, we take the reaction rate parameter values
$(g,k,\mu,\lambda)=(20,10,0.004,0.03)$ and initial condition
$(N_R,N_{R^*},N_A,N_{A^*})=(100,0,100,0)$. The evolution of the first two
moments of $N_{A^*}$ as computed from our approximate solution
Eq.~\ref{eq:phi0th1}, $\Omega-$expansion solution
Eqs.\ref{eq:ch1m},\ref{eq:ch2m}, and exact numerical results are shown in
Fig.~\ref{f:cas2m12}a,b. Unlike the practical implementation of the
$\Omega-$expansion, Eq.~\ref{eq:phi0th1} also directly gives the time
evolution of the full probability distribution for $A^*$ proteins. A time
slice of the probability distribution of $A^*$ is shown in
Fig.~\ref{f:cas2d}a. Overall, a remarkable agreement is achieved between the
approximate analytical results and exact numerical calculations. Also shown in
the figure is the distribution computed from the Langevin equation (dashed line), 
which is characterized by an average
noticeably larger than the exact average. Even though the magnitude of $R^*$
fluctuations is the same for the cascade parameters used in
Figs.~\ref{f:omgill},\ref{f:cas2m12},\ref{f:cas2d}, the $A^*$ fluctuations are
dramatically attenuated for the case demonstrated in Fig.~\ref{f:cas2m12}b and
\ref{f:cas2d}a, compared with Fig.~\ref{f:omgill}. In this case the $R \to
R^*$ reactions are much faster than the $A \to A^*$ reactions, thus, the $R
\to R^*$ noise is averaged out and only internal $A \to A^*$ noise remains. If
we think of this two-step cascade as an element of a longer signaling pathway,
the relatively large fluctuations of the upstream reaction ($R \to R^*$)
become attenuated downstream ($A \to A^*$). Thus, stacking of a slow
downstream reaction after a fast upstream reaction provides a general
mechanism for noise attenuation in biochemical signaling networks.

In the opposite limit of the reaction rate of $R^*$ being much slower than
that of $A$, we can take into account the $R^*$ dynamics by allowing
$\phi_{m,0}$ in Eq.~(\ref{eq:phi0th1}) to change slowly with time according to
Eq.~(\ref{eq:rdis}). Thus, we substituted
\begin{equation}
\phi_{m,0}=\exp(-(1-e^{-kt})g/k)(1-e^{-kt})^{m}(\frac{g}{k})^{m}/m!
\,. \label{eq:adsub}
\end{equation}
into Eq.~(\ref{eq:phi0th1}). This is a valid approximation in the limit of
slow $R^*$ dynamics. Using this substitution, we obtained an approximate
solution for $\Psi(x,y)$ as an infinite sum over the $R^*$ protein number
$m$. Since the coefficients of $x^m$ decay very fast with increasing $m$, we
can safely truncate it to a finite sum. For many reaction rates in this
regime, the obtained distribution is broad.

A set of reaction rates, which corresponds to a slow upstream reaction and a
fast downstream reaction $(g,k,\mu,\lambda)=(0.02,0.01,0.2,1.5)$, was used to
compare our analytical calculations with exact numerical results
(Fig.~\ref{f:cas2m12}c,d and Fig.~\ref{f:cas2d}b). The evolution of both the
first moment (Fig.~\ref{f:cas2m12}c) and the variance (Fig.~\ref{f:cas2m12}d)
obtained from our analytical treatment agrees well with the exact numerical
one, being more accurate than the $\Omega$-expansion. Furthermore, we used
Eq.~\ref{eq:newphi}, Eq.~\ref{eq:phi0th1}, and Eq.~\ref{eq:adsub} to compute
the full probability distribution at $t=200$ (Fig.~\ref{f:cas2d}b), which is
impossible to obtain analytically using the $\Omega$-expansion approach. All
the nuances of the complicated distribution are accurately captured by our
approximate analytic solution. The distribution computed from the Langevin
equation, on the other hand, which is shown as a dashed line
in Fig.~\ref{f:cas2d}b, is characterized by a single broad peak. The white
noise terms in the Langevin equation obviously smear out the peaks and, thus,
are not good models for the underlying stochastic dynamics. At long time
limits the distribution becomes uni-modal, but still wide (data not shown).

\begin{figure}[tbh] 
\centering
\subfigure[]{
\includegraphics[width=8cm]{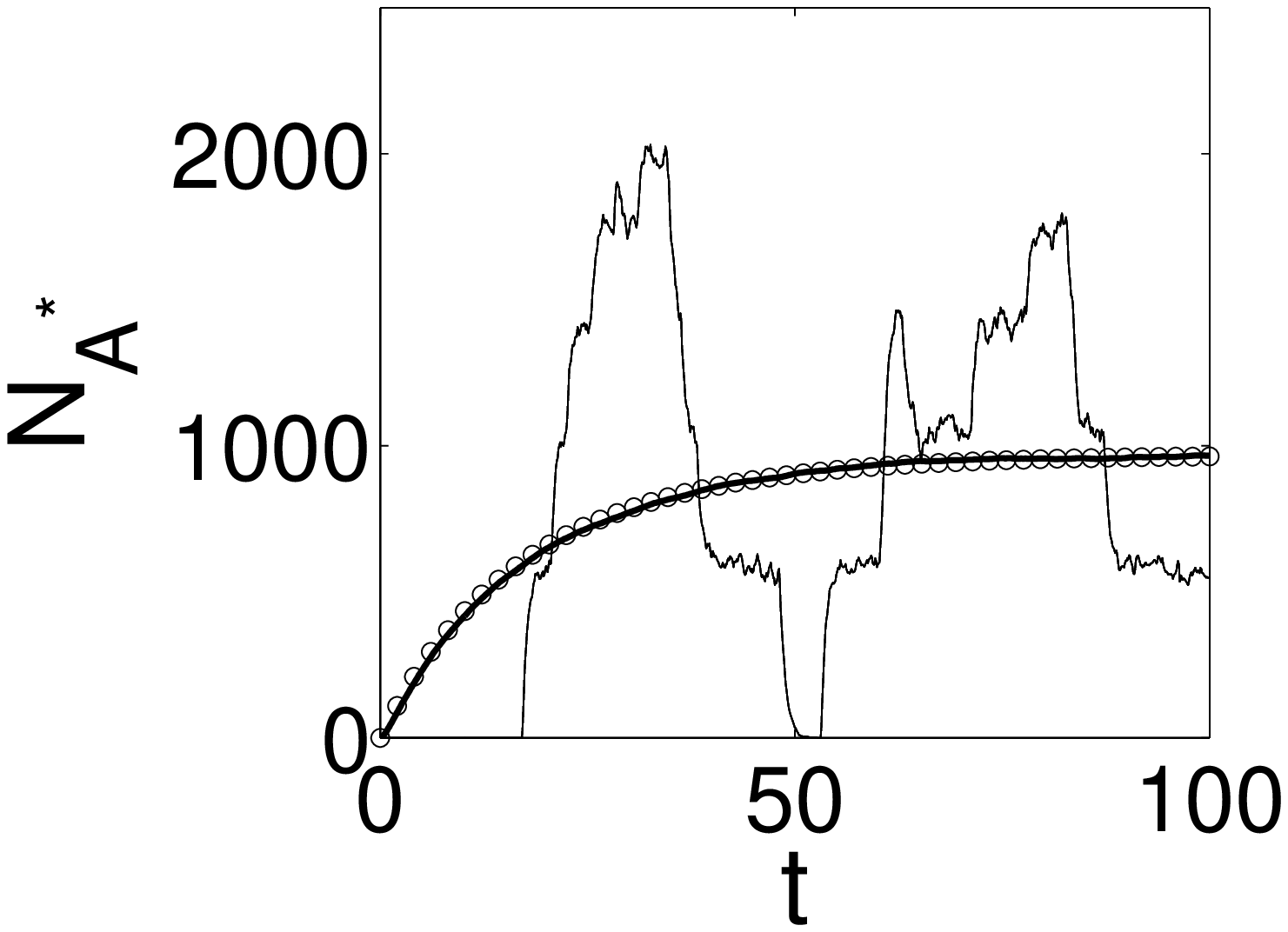}}
\subfigure[]{
\includegraphics[width=8cm]{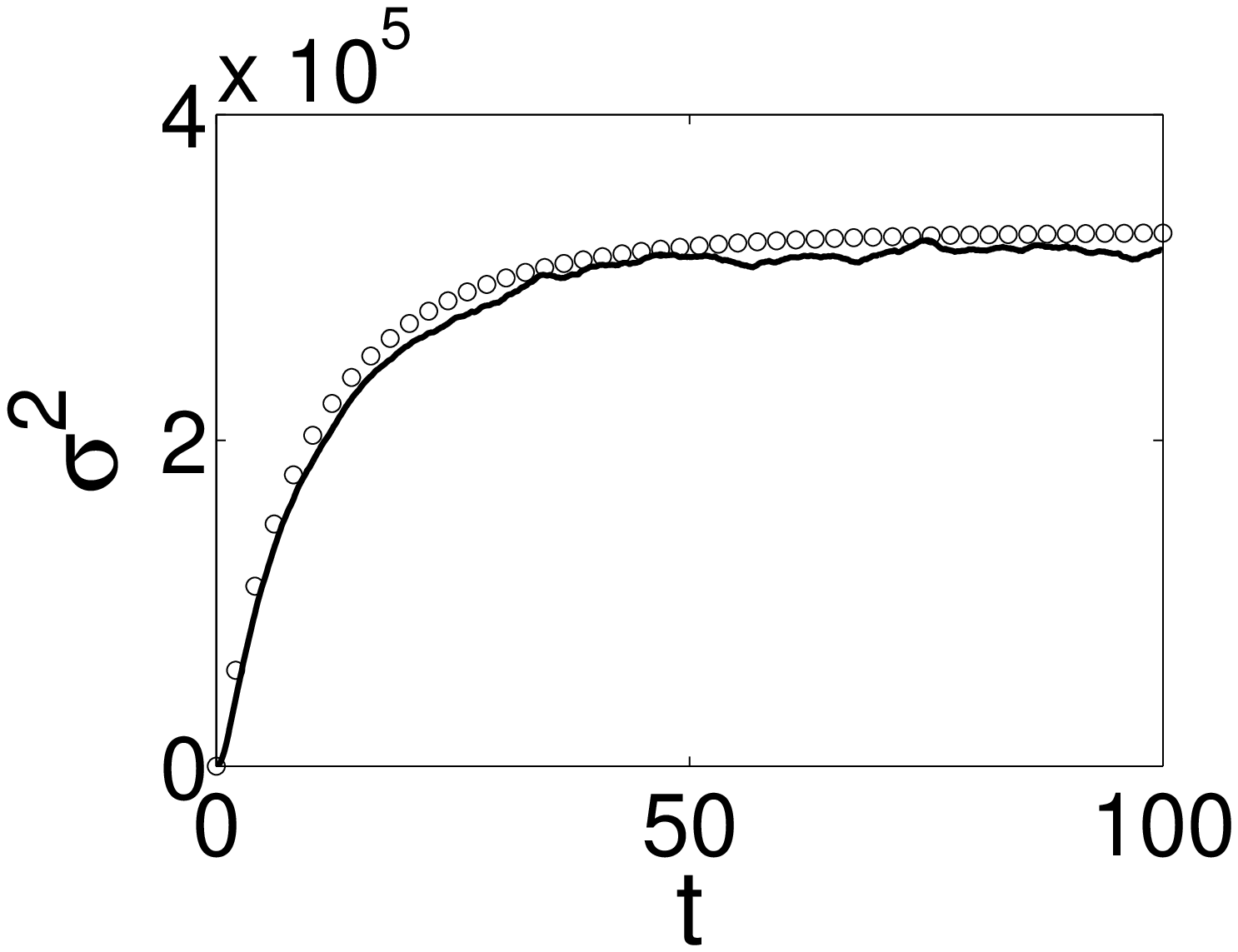}}
\caption{(a) Averages computed from Gillespie simulation ({\it solid line})
  and from the approximate solution Eq.~(\ref{eq:newphi}) ({\it circles}). One
  typical Gillespie trajectory is also shown ({\it thin solid line}); (b) the
  variance from Gillespie simulation ({\it solid line}) and from the
  approximation ({\it circles}). Initial condition
  $(N_R,N_{R^*},N_A,N_{A^*})=(100,0,5000,0)$ and parameter values
  $(g,k,\mu,\lambda)=(0.1,0.05,0.2,1.5)$.  }
\label{f:sslowm}
\end{figure}

The fluctuations in $A^*$ are much larger for a cascade where the upstream
reaction is slow and the downstream reaction is fast (Fig.~\ref{f:cas2m12}d
and Fig.~\ref{f:cas2d}b) compared with previously considered cascades
(Figs.~\ref{f:omgill}, \ref{f:cas2m12}b and \ref{f:cas2d}a). Thus, the noise
produced in the $R \to R^*$ reaction is retained and amplified by the next
enzymatic reaction. As opposed to the previously discussed case of
noise-attenuation, this cascade setup could be used to amplify the noise
downstream. The amplification and attenuation of noise have been extensively discussed and
experimentally tested in the linearized stochastic description of gene
regulatory networks\cite{th01in,kie01eff,oz02reg,bla03noi}. Our
current work emphasizes the role of discreteness in a nonlinear biochemical
reaction network, using directly the generating function formalism to obtain
analytical time-dependent probability distributions.

Strong fluctuations occur even when the number of downstream proteins is very
large. Fig. ~\ref{f:sslowm} demonstrates a striking example. Although the
average number of $A^*$ quickly reaches its steady-state value, which is near
$1000$, the fluctuation continues, such that a typical trajectory fluctuates
with a magnitude of the same order as the average (Fig.~\ref{f:sslowm}a). The
corresponding large variance, shown in Fig.~\ref{f:sslowm}b, confirms this
analysis. Thus, the fluctuations are much larger than expected from the usual
$\sqrt N$ argument, where N is the number of proteins. Our approximate
solution agrees very well with the exact solution (Fig.~\ref{f:sslowm}),
confirming the validity of our noise-amplification scheme to solve the master
equation (Eqs.~(\ref{eq:newphi}), (\ref{eq:phi0th1}) and
(\ref{eq:adsub})). The large $A^*$ fluctuations are induced by the upstream
$R^*$ fluctuations, as seen from a typical numerical trajectory shown in
Fig.\ref{f:sslowm}a. At steady state, the average death/birth timescale for
$R^*$ is $1/g=10$, thus, the $A^*$ trajectory exhibits a bursting behavior
with a time correlation of about $10$.

In the photoreception cascade, the rhodopsin (Rh) activation is controlled by
the incident photons which may arrive, one by one, within $1s$ or even longer
time interval in the single photon experiment\cite{sch95pht}. The deactivation
rate from Rh$^*$ to Rh is about $0.5 s^{-1}$. The activation rate of the
transducin\cite{bur01act} is around $120s^{-1}$ and the deactivation rate is
$100 s^{-1}$. According to the classification developed in this paper, the
initiating stage of the photoreception cascade belongs to the case of slow
$R^*$ reaction and fast $A$ reaction, thus, we expect an amplification of the
external light stimulation~\cite{sch95pht,rk98or,bur01act}.

Our approximation method in this section is based on the separation of time
scales for the first and the second reactions, an approach which is also the
starting point for many solution techniques in the literature. The elimination
of fast variables is among the most popular ones and is often used to
approximately solve Fokker-Planck or Langevin
equations~\cite{van92st,gar02han}. Various methods such as the projection
operator method~\cite{gar02han} and cumulant expansion method~\cite{van92st}
have been developed to treat the continuous cases. To the lowest order
approximation, the fast time scale is completely removed, while the equations
for the slow variables account for the fast variables either through their
corresponding averages~\cite{van92st} or through their stationary
distributions~\cite{gar02han}. Here we applied the essence of this idea to the
discrete jump process described by a master equation and obtained analytical
approximations for the evolving probability distribution function. Similar
consideration has been used to derive effective equations in gene
transcription regulation in the large particle limit~\cite{th01st}, but our
method applies to small particle numbers as well. In the first case considered
above (fast $A-A^*$ reaction), we used only the average of the fast variable.
In the second case (fast $R-R^*$ reaction), we first considered the
probability distribution evolution of the fast variable and, subsequently,
incorporated back the evolution of the slow variable. Thus, in both cases, we
explicitly included both the slow and fast time scales into the final
analytical expression.

\section{Solving the master equation with the hybrid smooth probability
  distribution method}
\label{sect:smooth}

When the probability distributions are relatively smooth, a new approximation
scheme may be employed to integrate the generating function equation
\ref{eq:m4dx}. To show basic idea of the method, we first implement it for the
2-step cascade discussed previously (Fig. \ref{f:react}). Then, to demonstrate
the potential for generalization, we apply this method to another enzymatic
cascade, where the first step is a self-dimerization of receptor $R$ instead
of a simple activation.

\subsection{2-step signal transduction cascade}
To treat analytically the stochastic signaling dynamics in a wider regime of
parameters (for example, when upstream and downstream reactions have
comparable rates), the neglected coupling terms in Eq.\ref{eq:newexp} may be
taken into account in a more systematic way.  In this section, we will
reconsider the cross terms between different $m$'s and treat them in a
different way. Here, we take advantage of the known $R^*$ distribution from
Eq.~\ref{eq:ysolexp} and write down the following expansion for $\Psi(x,y)$,
\begin{equation}
\Psi(x,y)=\sum_{m=0}^{\infty}\exp(-\frac{g}{k}(1-e^{-kt}))
(\frac{g}{k})^m \frac{x^m}{m!}\phi_m(y)
\,, \label{eq:appsi}
\end{equation}
where $\phi_m(y)$ is a time-dependent function and, according to
Eq.~(\ref{eq:ysolexp}), we know that $\phi_m(1)=(1-e^{-kt})^m$. Substituting
this form of expansion of $\Psi$ into Eq.~(\ref{eq:m4dx}) and comparing the
coefficient of $x^m$ on both sides of the equation, we have
\begin{widetext}
\begin{equation}
\frac{\partial \phi_m}{\partial t}=(1-y)(\mu m \frac{\partial}{\partial y}-\lambda N
+\lambda y\frac{\partial}{\partial y})\phi_m
+km(\phi_{m-1}-\phi_m)+g(\phi_{m+1}-\phi_m)+ge^{-kt}\phi_m
\,.\label{eq:appexp}
\end{equation}
\end{widetext}
This equation describes the time evolution of $\phi_m(y)$, which is coupled to
the neighboring functions $\phi_{m-1}$ and $\phi_{m+1}$. In previous
discussions, we neglected these couplings first and only later incorporated
them back in an effective way, justified under certain conditions. Here, we
present an approach which directly takes into account these couplings in an
approximate manner. When the probability distribution of $R^*$is smooth, the
following approximation,
\begin{equation}
\frac{\partial \phi_m}{\partial m} \approx \phi_{m+1}-\phi_m 
\approx \phi_m-\phi_{m-1}
\label{eq:diffapp}\,,
\end{equation}
may be used to uncouple the PDEs in Eq.~(\ref{eq:appexp}):
\begin{widetext}
\begin{equation}
\frac{\partial \phi_m}{\partial t}=(1-y)(\mu m \frac{\partial}{\partial y}-\lambda N
+\lambda y\frac{\partial}{\partial y})\phi_m
+(g-km)\frac{\partial \phi_m}{\partial m}+ge^{-kt}\phi_m
\,,\label{eq:newexpap}
\end{equation}
\end{widetext}
We solved the resulting PDEs by the method of
characteristics. Eq.~(\ref{eq:diffapp}) is satisfactory when the profile of
$\phi_m$ between $m$ and $m+1$ can be reasonably approximated by a straight
line segment. This works well if either of the two conditions holds: (1) as
mentioned above, the distribution profile is smooth so that the higher-order
derivatives can be ignored, or (2) the number of $R^*$ is large such that an
increment by one particle may be treated as small. In addition, it is possible
to improve this solution by making a higher-order approximation to the
difference in Eq.~(\ref{eq:diffapp}). For example, an exact expression can be
obtained through the Kramers-Moyal expansion\cite{gar02han}
\[
\phi_{m+1}-\phi_m=\sum_l \frac{1}{l!}\frac{\partial^l}{\partial m^l}\phi_m
\,,
\]
where $l$ runs from $1$ to $\infty$. In the standard derivation of the
Fokker-Planck equation\cite{van92st,gar02han}, terms up to the second order
($l=2$) are retained. In some sense, our ``smooth distribution'' method is a
hybrid distribution function -- generating function scheme, where in the yet
to be determined functions $\phi_m(y)$, subscript $m$ is related to the
$R^\star$ particle number (distribution for $m$), while $y$ is a formal
variable related to the generating function for the $A^\star$ particle number.
However, solving equations containing higher-order derivative terms quickly
becomes cumbersome. Here, we only keep the first order term, which allows
Eq.~(\ref{eq:newexpap}) to be solved with the following set of characteristic
equations
\begin{eqnarray}
\dot{y} &=& (y-1)(\mu m+\lambda y) \nonumber \\
\dot{m} &=& mk-g \nonumber \\
\dot{\phi_m} &=& \lambda N (y-1)\phi_m+g e^{-kt} \phi_m 
\,. \label{eq:newch}
\end{eqnarray}
The first two equations in Eq.~(\ref{eq:newch}) define the characteristic
curves and the third equation shows how $\phi_m$ changes along this curve. The
dynamics on each curve is self-contained and independent of each other, which
is the consequence of neglecting higher order terms in Eq.~(\ref{eq:diffapp}).
Eq.~(\ref{eq:newch}) were exactly solved, resulting in
\begin{equation}
\phi_m(y)=\phi_m^{(0)}(z_0)(1+\frac{\lambda p}{p^\prime}(y-1))^N 
\exp[\frac{g}{k}(1-e^{-kt})]
\,,\label{eq:newchsl}
\end{equation}
where
\begin{eqnarray*}
\phi_m^{(0)} &=& \sum_{n=0}^{N}a_{nm}(\frac{1}{z_0}+1)^n  \\
z_0 &=& \frac{p^\prime}{y-1}+\lambda p \\
p(t) &=& \int_0^t e^{I(s)} ds \\
I(t) &=& (\lambda+\frac{\mu g}{k})t+\frac{\mu}{k}(m_0-\frac{g}{k})(e^{kt}-1)
\,.
\end{eqnarray*}
$a_{nm}$ is the initial probability of having $m$ $R^*$'s and $n$ $A$'s and
$m_0$ is an intermediate variable which after all the integrations and
differentiations are done will be replaced by $m$ by the following
substitution,
\begin{equation}
m_0-\frac{g}{k}=(m-\frac{g}{k})e^{-kt}
\,.\label{eq:msub}
\end{equation}
In general, the integration to obtain $p(t)$ can not be carried out in a
closed form, thus, approximate or numerical treatment is needed. However, the
analytic structure of the overall solution is transparent. The last
exponential factor in Eq.~(\ref{eq:newchsl}) describes the $R^*$ reaction and
the first two factors describe the $A \to A^*$ reaction. The full generating
function $\Psi(x,y)$ is given by Eq.~(\ref{eq:appsi}) with $\phi_m(y)$ given
by Eq.~(\ref{eq:newchsl}). Once $\Psi(x,y)$ is known, all the statistical
quantities are easily computable.

As an example, we consider the following initial distribution
\begin{equation}
\phi_m^{(0)}=e^{-m}(\frac{1}{z_0}+1)^N
\,,\label{eq:newexpinit}
\end{equation}
which corresponds to starting the cascade dynamics with $N$ $A$'s and a small
number of $R^*$, distributed exponentially. The corresponding solution is
given by
\begin{equation}
\phi_m(y)=(1+\frac{\lambda p +1}{p^\prime}(y-1))^N \exp(-m e^{-kt})
\,,\label{eq:newexpsol}
\end{equation}
where
\begin{widetext}
\begin{eqnarray}
p^\prime (t) &=& \exp ((\lambda+\frac{\mu g}{k})t+\frac{\mu}{k} (m-\frac{g}{k})
(1-e^{-kt})) \nonumber \\
p(t) &=& \int_0^t \exp ((\lambda+\frac{\mu g}{k})s+\frac{\mu}{k} (m-\frac{g}{k})
(e^{k(s-t)}-e^{-kt})) \,ds
\,.\label{eq:newexpp}
\end{eqnarray}
\end{widetext}
When $x=1$, if the following approximation is used
\begin{equation}
\phi_m=\exp(-m e^{-kt}) \approx (1-e^{-kt})^m
\,, \label{eq:fapp}
\end{equation}
then the $R^*$ distribution is recovered. Furthermore,
Eq.~(\ref{eq:newexpsol}) with Eq.~(\ref{eq:fapp}) substituted in conserves the
total probability, {\em i.e.}, $\Psi(1,1)=1$. Thus, we use Eq.~(\ref{eq:fapp})
in the calculations described below.

\begin{figure}[tbh] 
\centering
\subfigure[The distributions at $t=6$]{
\includegraphics[width=8cm]{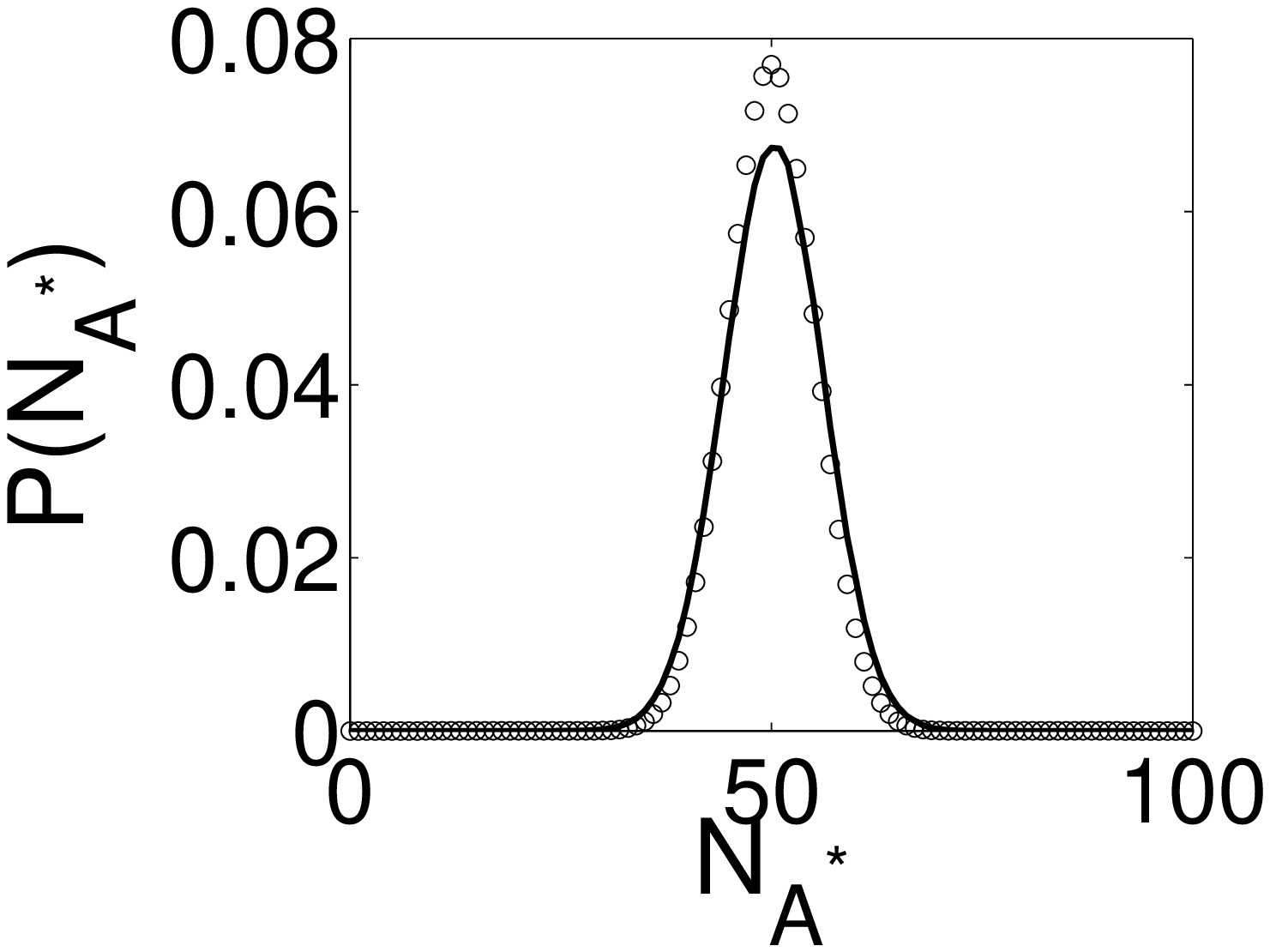}}
\subfigure[The distributions at $t=60$]{
\includegraphics[width=8cm]{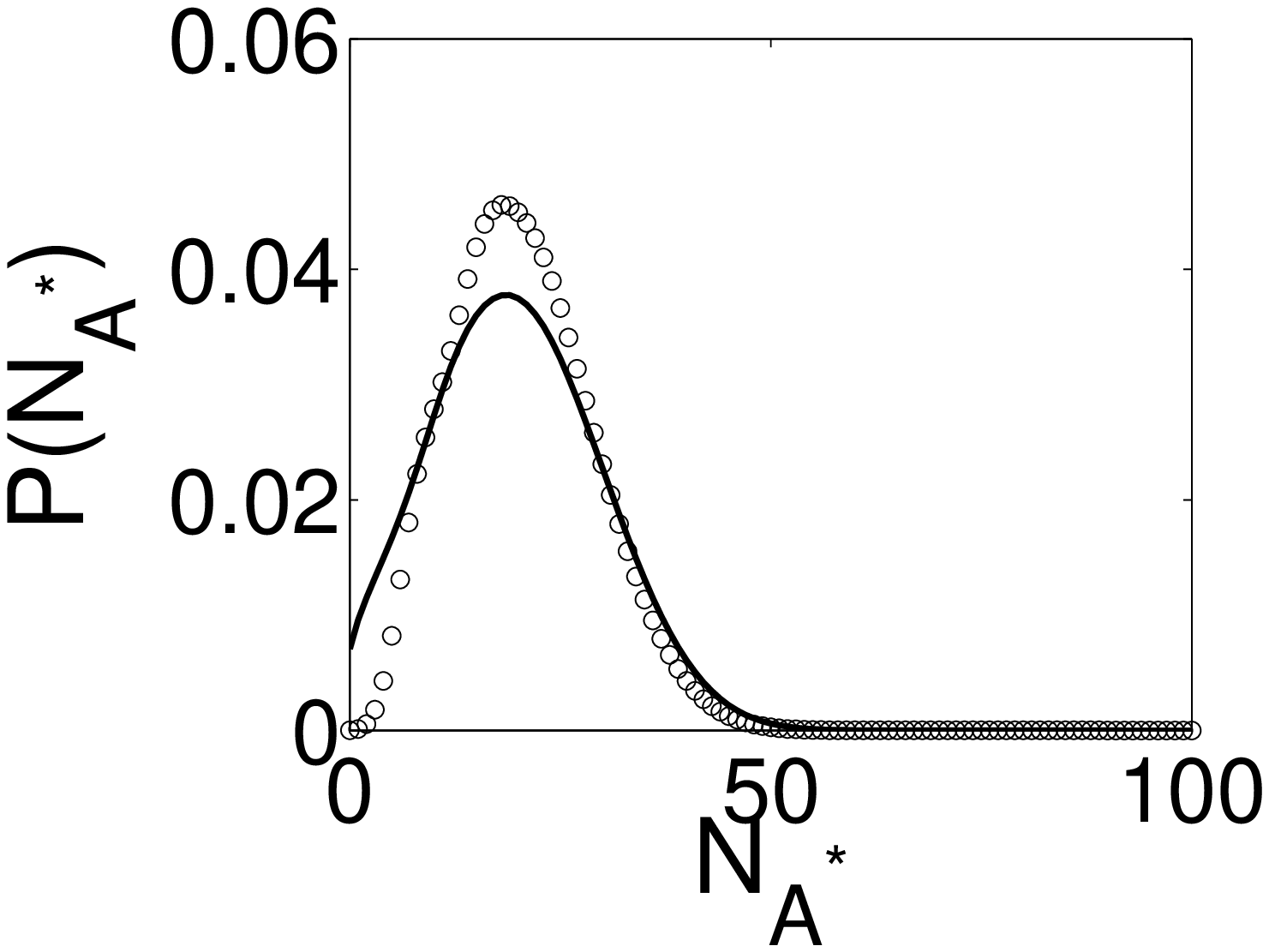}}
\caption{The probability distributions of $A^*$ at $t=60$ computed from the
  approximate solution Eq.~(\ref{eq:newexpsol}) ({\it circles}) and the exact
  solution Eq.~(\ref{eq:m1}) ({\it solid line}) with initial condition
  $(N_R,N_{R^*},N_A,N_{A^*})=(100,0,100,0)$ and parameter values: (a)
  $(g\,,k\,,\mu\,,\lambda)=(10\,,1\,,0.01\,,0.1)$; (b)
  $(g\,,k\,,\mu\,,\lambda)=(0.2\,,0.1\,,0.02\,,0.2)$.}
\label{f:bendisd}
\end{figure}

To gauge the effectiveness of the approximate analytical solution, we take
$N=100$ and the $R^*$ distribution truncated at $m=30$. Firstly, we evaluate
Eq.~(\ref{eq:newexpsol}) with
$(g\,,k\,,\mu\,,\lambda)=(10\,,1\,,0.01\,,0.1)$. $A^*$ distribution computed
from Eq.~(\ref{eq:appsi}) at $t=60$ matches quite well with the exact solution
as shown in Fig.~\ref{f:bendisd}a. The approximate distribution is a little
narrower than the exact one due to the omission of the higher-order derivative
terms in Eq.~(\ref{eq:diffapp}). Secondly, we carried out similar calculations
with $(g\,,k\,,\mu\,,\lambda)=(0.2\,,0.1\,,0.02\,,0.2)$, as shown in
Fig.~\ref{f:bendisd}b. Although the distribution at $t=60$ from our
approximation agrees quite well with the exact solution for large $N_{A^*}$,
near the left boundary there is a clear discrepancy, due to the non-smoothness
of the distribution at the minimum particle number, since the number of $A^*$
cannot be negative.

\subsection{Receptor self-dimerization introduces additional nonlinearity}

\begin{figure}[tbh]
\centering
\includegraphics[width=12cm]{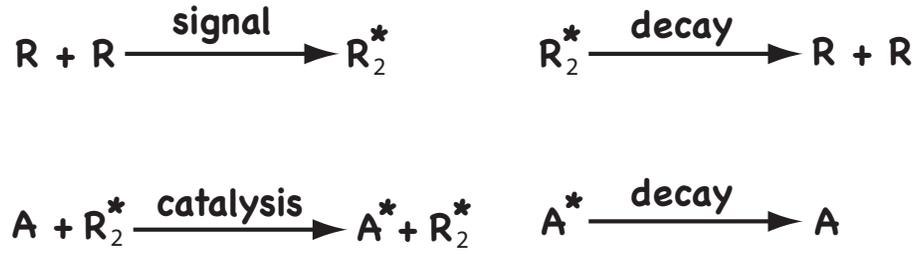}
\caption{An inactive receptor $R$, when activated by dimerization, activates
  downstream protein $A$.}
\noindent 
\label{f:reactd}
\end{figure}

Even if the exact solution is not available for the first reaction, we can
still apply the above approximation as long as the distribution is
smooth. Consider the dimerization reaction shown in
Fig. \ref{f:reactd}. Compared to the previously discussed 2-step cascade
(Fig. \ref{f:react}), the first reaction is replaced by a self-dimerization
process with rate $g$. This dimerization activation is quite common in signal
transduction and gene regulatory
networks~\cite{jas04fl,bun03fl,hay04lin}. Although it is possible to derive an
analytical solution for the isolated first step in the cascade, i.e. the
self-dimerization process~\cite{nic77self,rei98st}, the expression is in a
series form and will not be used in our approximation scheme.

Again, let $P(m,n)$ denote the probability of having $m$ $R_2^*$'s and $n$
$A$'s. The master equation is then,
\begin{eqnarray}
\frac{dP}{dt}(m,n)&=& \frac{g}{2}\left( (M-2m+2)(M-2m+1)P(m-1,n)
-(M-2m)(M-2m-1)P(m,n)\right) \nonumber \\
&&+k\left( (m+1)P(m+1,n)-mP(m,n) \right)+\mu\left((n+1)mP(m,n+1)
-mnP(m,n) \right) \nonumber \\
&&+\lambda\left((N-n+1)P(m,n-1)-(N-n)P(m,n) \right)
\,, \label{eq:dimmeq}
\end{eqnarray}
where the first term describes the dimerization reaction. The corresponding
generating function $\Psi(x,y)$ satisfies,
\begin{eqnarray}
\frac{\partial \Psi}{\partial t}&=& \frac{g}{2}(x-1)\left(M(M-1)+4x^2\frac{\partial^2}{\partial x^2}
-2(2M-3)x\frac{\partial}{\partial x} \right) \Psi \nonumber \\
& &+k(1-x)\frac{\partial \Psi}{\partial x}+\mu(1-y)x\frac{\partial^2 \Psi}{\partial x \partial y}
+\lambda (y-1)(N-y\frac{\partial}{\partial y})\Psi
\,. \label{eq:dimpde}
\end{eqnarray}
Note that a new second-order derivative, $\partial^2 \Psi/\partial x^2$,
appears, compared with the previously considered generating function PDE
(Eq. \eqref{eq:m4dx}).  If we expand the generating function in the form of
Eq.~(\ref{eq:newphi}), a series of PDEs for $\phi_m$'s are derived. Similar to
the simple 2-step cascade, after taking the continuous limit approximation, we
get
\begin{eqnarray}
\frac{\partial \phi_m}{\partial t} & \approx & (1-y)(\mu m \frac{\partial}{\partial y}
-\lambda N+\lambda y \frac{\partial}{\partial y}  )\phi_m  \nonumber \\
& & +\frac{\partial}{\partial m}\left(km-\frac{g}{2}(M(M-1)+4m(m-1)-2(2M-3)m) \right)\phi_m
\,, \label{eq:dimapp}
\end{eqnarray}
where the total probability is conserved since
\begin{equation*}
\frac{d}{dt}\int_{-\infty}^{\infty} dm \, \phi_m(t,1)=0
\,.
\end{equation*}
Eq.~(\ref{eq:dimapp}) are also readily solved analytically by the characteristic method.
\begin{figure}[tbh] 
\centering
\subfigure[The distribution at t=3]{
\includegraphics[width=8.cm]{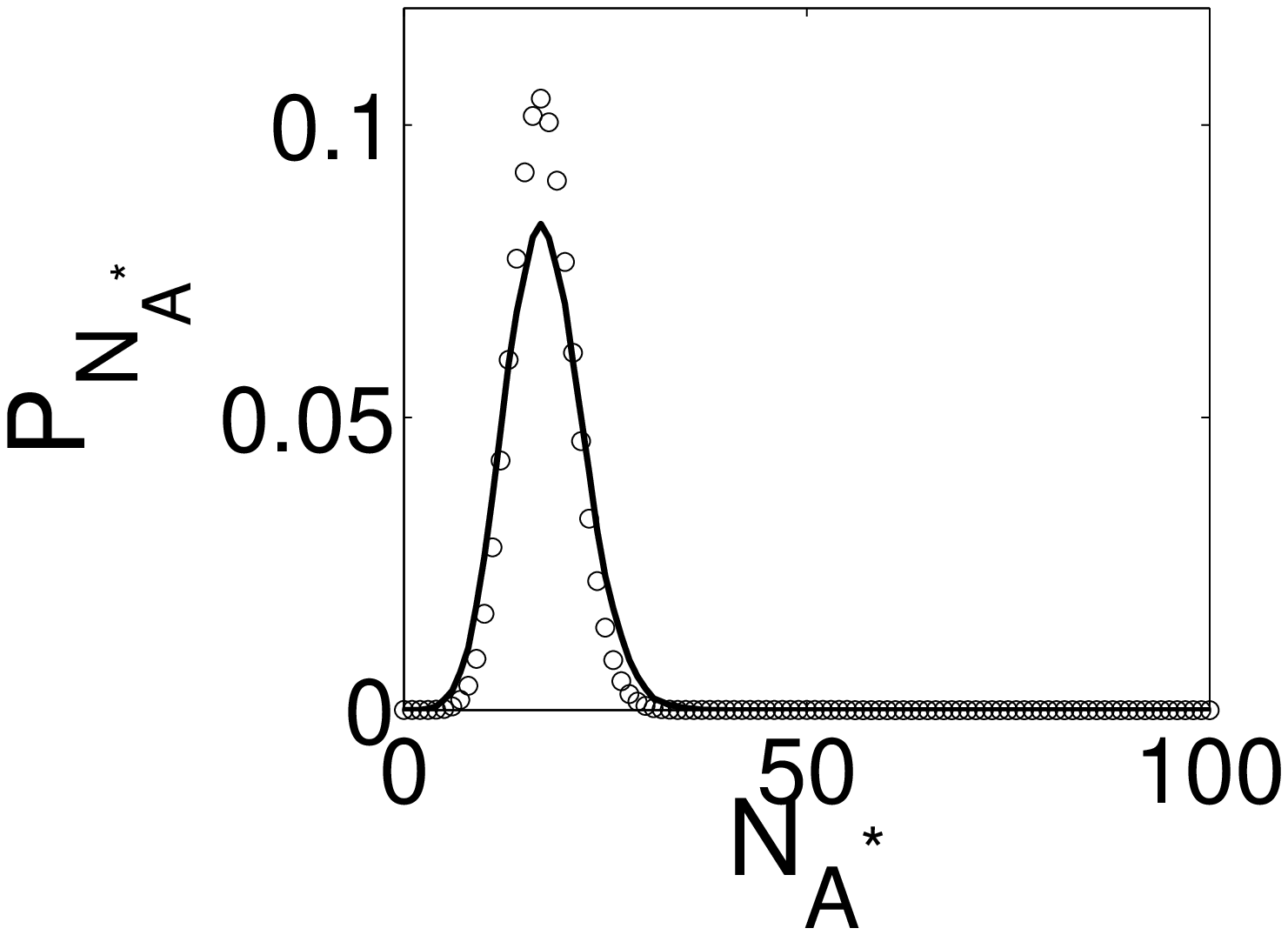}}
\subfigure[The distribution at t=60]{
\includegraphics[width=8.cm]{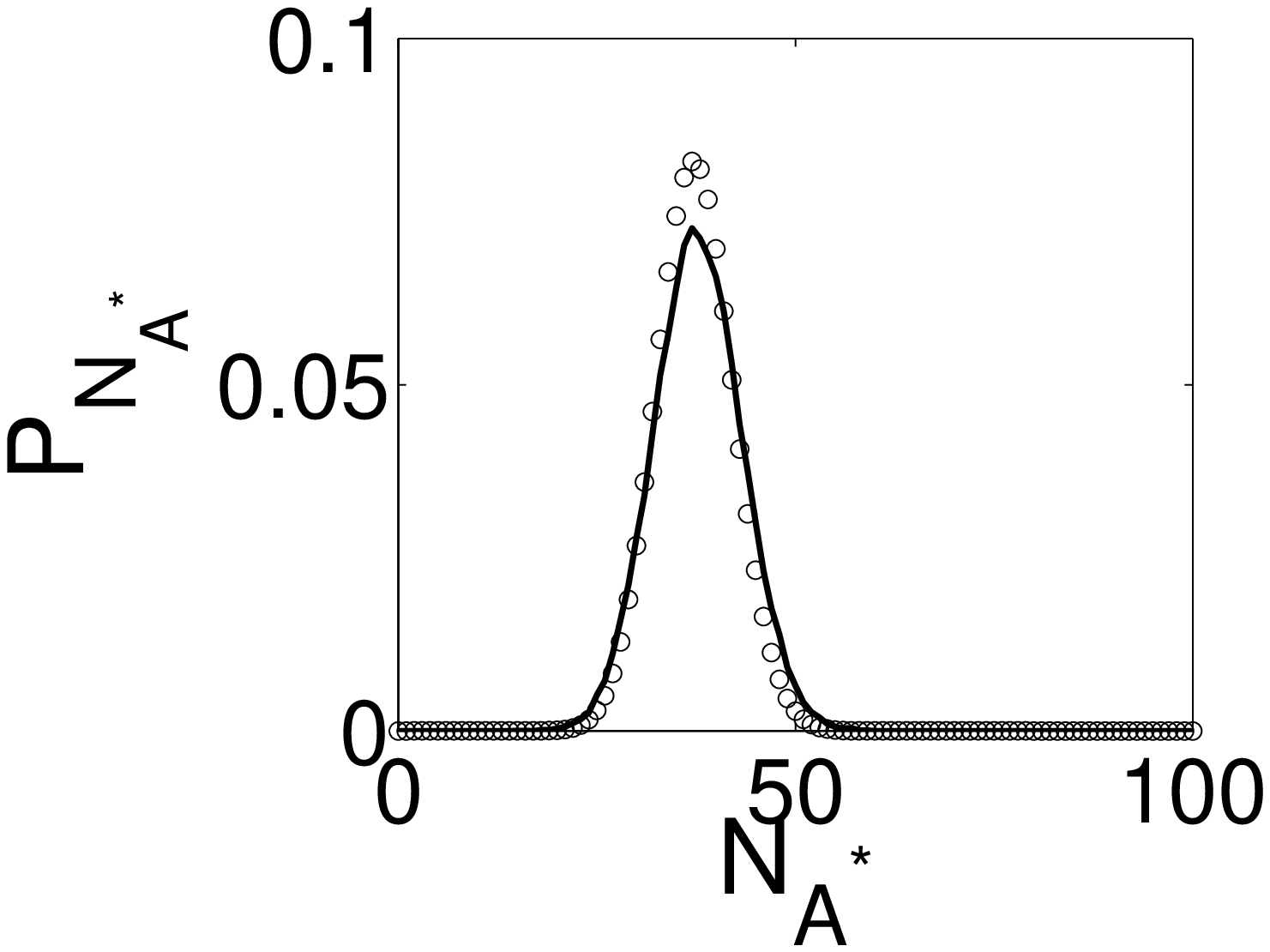}}
\caption{Probability distributions of $A^*$ computed from the approximate
  solution Eq.~\ref{eq:dimapp} ({\it circles}) and Gillespie simulation (solid
  line) averaged over $70898$ realizations.  Initially $(N_A,N_{A^*})=(100,0)$
  and $N_{R_2^*}$ approximately assumes a Gaussian distribution
  $\exp(-(m-2)^2)/\sqrt{\pi}$. We use $N_R+2N_{R_2^*}=20$ and choose parameter
  values $(g,k,\mu,\lambda)=(0.02,0.5,0.02,0.15)$. (a) The distribution
  $P(N_{A^*})$ at $t=3$. (b) The distribution $P(N_{A^*})$ at $t=60$.  }
\label{f:dimapp}
\end{figure}

The approximate distributions compared with those computed from Gillespie
simulation are displayed in Fig. \ref{f:dimapp} at an early time $t=3$ and a
later time $t=60$. They agree quite well. The approximate and exact
distributions at other times are in good agreement as well (data not shown).

Overall, the presented method may be used to obtain the long time evolution of
the stochastic signaling dynamics. We designed the hybrid scheme to treat the
second order derivative terms in the generating function equation with the
characteristics method.  When the probability distribution profiles of all
species are relatively smooth the method gives quantitatively accurate
results. Although, this condition is most commonly satisfied when protein
numbers are large, it is also applies to systems with smaller protein numbers
with certain constraints on the relative reactions rates. Generalizing this
method to treat larger cascades is also straightforward. If all the reaction
nodes in the network are linear, then there is no need to expand the
generating function equation up to second order derivative terms, and a direct
application of the characteristics method would solve the generating function
PDE.  If a small number of binary reaction nodes are present, the hybrid
scheme can be applied most effectively, allowing to obtain approximate
solutions in a manner similar to the examples that we have already
considered. For cascades with many binary reactions, a straightforward
application of the method may becomes unpractical since a summation over many
indices is required, which would be computationally expensive. However, this
difficulty may be overcome by either linearizing the nonessential binary terms
or by approximating the summations with integrals.

\section{Conclusions}

Cells live in a fluctuating environment in which signals and noise keep
bombarding the cell receptors\cite{st02gomp,bar98bac,sm05inf}. Noisy signals
propagate inside the cell via microscopic chemical reaction events. The
external noise interferes with the internal biochemical network noise
originating from the underlying fundamental randomness of these chemical
reactions. Cells have evolved to adapt to or even exploit the seemingly
deleterious effect of fluctuations on signaling dynamics within a mesoscopic
size object. Thus, it is important to develop a qualitative picture, based on
mathematical modeling of stochastic chemical kinetics, of how signaling
networks process noisy signals. In this paper we studied the stochastic signal
transduction by a simple 2-step signaling cascade using the master equation to
describe stochastic reaction events. In agreement with previous studies, we
found that when particle numbers are large, chemical kinetic equations provide
an accurate description.  However, when the number of proteins becomes small,
a large variability among individual trajectories results, necessitating the
stochastic chemical kinetics approach. If fluctuations are small, the commonly
used linear noise approximation works well, the probability distribution being
centered around the deterministic trajectory. But when fluctuations are large,
for example, at the initiating (burst) phase of a signal transduction cascade,
the linear noise approximation breaks down and more powerful analytical
treatment of the master equation becomes necessary. In the small protein
number regime, chemical Langevin equation does not work properly as well since
the continuous assumption breaks down and molecular discreteness sets in the
dynamics.

Without assuming that noise is small, we directly treated the master equation
with a generating function approach. Although the resulting PDE could not be
solved exactly, we found a number of perturbative schemes, that allow us to
obtain approximate analytic solutions for the generating function which is
used to obtain the time evolution of the full probability distribution for all
proteins. Using the analytic solution, we recovered the general mechanism for
attenuating or amplifying noise in a signaling pathway with nonlinear reaction
events: if the upstream reactions are fast and the downstream reactions slow,
then upstream noise becomes attenuated. Conversely, if the upstream reactions
are slow and the downstream reactions fast, then the upstream noise becomes
amplified. Thus, controlling various node timescales by regulating reaction
rate constants would lead to enhancing of the signaling cascade sensitivity
and reliability by suppressing uncorrelated noise while still amplifying weak
signals. This mechanism may be used by cells to draw useful information from a
noisy environment. Furthermore, under certain conditions, the burst phase
noise may induce macroscopic system-wide disturbance in the downstream
signaling network.

The approximation based on characteristics, presented in Section
\ref{sect:smooth}, can be straightforwardly generalized to a longer cascade,
with the restriction that the protein number distributions are smooth. Yet
another powerful technique to solve the master equation for more complex
cascades is based on the variational principle~\cite{eyink96act}. The analytic
solutions developed in this paper serve as a starting point for developing
high quality time-dependent basis sets for the variational approach. A good
basis set should capture the essential part of the system dynamics so as to
make the subsequent calculations simple and effective, which will be discussed
in more detail elsewhere. For larger signaling pathways, especially when
embedded in space, the commonly used numerical stochastic simulations will
face severe computational bottleneck. In this work we have taken a different
approach, based on analytically solving the master equation describing
stochastic chemical kinetics, to achieve, in an efficient manner, qualitative
and quantitative insights into stochastic signaling by biochemical reaction
networks. In an ongoing work we are generalizing the techniques presented in
this work to investigate the interplay of the noise amplification and
attenuation with the complex dynamics generated from feedback loops in larger
biochemical reaction networks.
 
\appendix
\section*{Appendix A: The $\Omega$-expansion of the 2-step cascade master
  equation}

If the fluctuations are relatively small, van Kampen's $\Omega$-expansion may
be used to account for their effect on signaling dynamics\cite{van92st}. In
this case it is more convenient to rewrite the master equation, given by
Eq.~\ref{eq:m1}, in an alternative form\cite{van92st}, where the dependence of
reaction rates on the cell size, $\Omega$, is explicitly emphasized:
\begin{equation}
\dot{P}=\lambda(E_n^{-1}-1)(N-n)P+\frac{\mu}{\Omega}(E_n^{+1}-1)mnP
+k(E_n^{+1}-1)mP+g\Omega(E_n^{-1}-1) P
\,,\label{eq:mch}
\end{equation}
where $E_n^{+1}f(n)=f(n+1)$ and $E_n^{-1}f(n)=f(n-1)$ are step-up and
step-down operators\cite{van92st}. Although $\Omega$ is equal to 1 in our
units, it is still written out to be used later as an expansion parameter. If
the protein numbers are large enough such that the deterministic evolution
serves as a good starting point, then small fluctuations are well described by
the linear noise approximation. In particular, the variables are changed to
emphasize the fluctuations around the deterministic orbits:
\begin{eqnarray}
m &=& \Omega \phi(t)+\Omega^{1/2} \xi      \,, \nonumber \\
n &=& \Omega \psi(t)+\Omega^{1/2} \eta     \,, \nonumber\\
P(m,n,t) &=& \Pi(\xi,\eta,t)     \,, \nonumber \\
N &=& \bar{N} \Omega
\,. \label{eq:chchv}
\end{eqnarray}
where $\phi(t)$ traces the deterministic path for $N_{R^*}$ and $\psi(t)$ for
$N_A$. The new random variables are $\xi$ and $\eta$, that describe
fluctuations around the average path. While averages determined from the
dominant paths are proportional to $\Omega$, fluctuations around these
averages are only proportional to $\Omega^{1/2}$. Since $\phi(t)$ and
$\psi(t)$ can be easily found by solving the chemical kinetic equation, we
need to obtain an evolution equation for the probability distributions of
$\xi$ and $\eta$, $\Pi(\xi,\eta,t)$. Thus, we substitute Eq.~\ref{eq:chchv}
into Eq.~\ref{eq:mch}, using also the following expression for any analytic
function $f(n)$:
\[
(E_n^{\pm 1}-1)f(n) = (\pm \frac{\partial }{\partial n}+\frac{1}{2}
\frac{\partial^2}{\partial n^2}+\dots)f(n) \,.
\]
Eq.~\ref{eq:mch} results in
\begin{eqnarray}
& & \frac{\partial \Pi}{\partial t}
-\Omega^{1/2}\frac{d \phi}{dt}\frac{\partial \Pi}{\partial \xi}-
\Omega^{1/2}\frac{d \psi}{dt}\frac{\partial \Pi}{\partial \eta}
=\lambda(-\Omega^{-1/2}\frac{\partial}{\partial \eta}+\frac{1}{2}\Omega^{-1}
\frac{\partial^2}{\partial \eta^2})(\Omega(\bar{N}-\Psi)-\Omega^{1/2}\eta)\Pi
\nonumber \\
& & +\frac{\mu}{\Omega}(\Omega^{-1/2}\frac{\partial}{\partial \eta}
+\frac{1}{2}\Omega^{-1}\frac{\partial^2}{\partial \eta^2})(\Omega \phi
+\Omega^{1/2}\xi)(\Omega \psi+\Omega^{1/2}\eta)\Pi
+k(\Omega^{-1/2}\frac{\partial}{\partial \xi}+\frac{1}{2}\Omega^{-1}
\frac{\partial^2}{\partial \xi^2}) \nonumber \\
& &(\Omega \phi+\Omega^{1/2} \xi)\Pi
+g(-\Omega^{-1/2}\frac{\partial}{\partial \xi}+\frac{1}{2}\Omega^{-1}
\frac{\partial^2}{\partial \xi^2})\Omega \Pi+\cdots
\,.\label{eq:chchexp}
\end{eqnarray}
We next collect terms that are of the same order in $\Omega$. The largest
$\Omega^{1/2}$ terms give:
\[
-\frac{d \phi}{dt}\frac{\partial \Pi}{\partial \xi}-
\frac{d \psi}{dt}\frac{\partial \Pi}{\partial \eta}=
-\lambda (\bar{N}-\psi)\frac{\partial \Pi}{\partial \eta}+\mu \phi \psi 
\frac{\partial \Pi}{\partial \eta}+k\phi\frac{\partial \Pi}{\partial \xi}
-g\frac{\partial \Pi}{\partial \xi}
\,,
\]
which is satisfied as we choose the dynamics of $\phi$ and $\psi$ to follow
chemical kinetics
\begin{eqnarray}
\frac{d \phi}{dt} &=& g-k\phi \,, \nonumber \\
\frac{d \psi}{dt} &=& \lambda(\bar{N}-\psi)-\mu \phi \psi
\,. \label{eq:chkn2}
\end{eqnarray}

At the $\Omega^0$ order, we obtain 
\begin{equation}
\frac{\partial \Pi}{\partial t}=\lambda \frac{\partial}{\partial \eta}(\eta \Pi)
+\mu\frac{\partial}{\partial \eta}(\xi \psi+\eta \phi)\Pi+k\frac{\partial}
{\partial \xi}(\xi \Pi)+\frac{1}{2}(\lambda(\bar{N}-\psi)+\mu \phi\psi)
\frac{\partial^2 \Pi}{\partial \eta^2}+\frac{1}{2}(k\phi+g)
\frac{\partial^2 \Pi}{\partial \xi^2}
\,, \label{eq:chfk}
\end{equation}
which is the familiar Fokker-Planck equation, derived in a systematic
way. Eq.~\ref{eq:chfk} is the linear noise approximation to the full
stochastic dynamics, which is valid when the path determined by
Eq.~\ref{eq:chkn2} is stable\cite{van92st}. Eq.~\ref{eq:chfk} is a linear PDE
that has to be solved numerically. However, it is possible to derive a closed
set of ODEs to describe time evolutions of the moments up to any order. For
example, the averages (first moments) satisfy
\begin{eqnarray}
\frac{d\langle\eta\rangle}{dt} &=& -(\lambda+\mu\phi)\langle\eta\rangle-\mu\psi\langle\xi\rangle \nonumber \\
\frac{d\langle\xi\rangle}{dt} &=& -k\langle\xi\rangle
\,,\label{eq:ch1m}
\end{eqnarray}
which are just the linearized chemical kinetics equations. Note that if
initial values of $\langle\eta\rangle$ and $\langle\xi\rangle$ are taken to be
zero, then they remain zero for all later times, consistent with the physical
significance of Eq.~\ref{eq:chkn2} which describes the evolution of averages
of protein numbers. This also suggests that application of Eq.~\ref{eq:chfk}
is based on the assumption of the validity of averaged chemical kinetics
equations. Next, we consider three second moments, satisfying
\begin{eqnarray}
\frac{d\langle\eta^2\rangle}{dt} &=&  -2(\lambda+\mu\phi)\langle\eta^2\rangle-2\mu\psi\langle\eta \xi\rangle
+\lambda(\bar{N}-\psi)+\mu \phi \psi \nonumber \\
\frac{d\langle\eta\xi\rangle}{dt} &=& -\lambda\langle\eta \xi\rangle-\mu\psi\langle\xi^2\rangle-\mu\phi\langle\eta \xi\rangle
-k\langle\eta\xi\rangle   \nonumber \\
\frac{d\langle\xi^2\rangle}{dt} &=& -2k\langle\xi^2\rangle+k\phi+g  
\,,\label{eq:ch2m}
\end{eqnarray}
to be solved simultaneously with Eq.~\ref{eq:chkn2}. In the current case,
Eq.~\ref{eq:chkn2},\ref{eq:ch1m},\ref{eq:ch2m} may be solved analytically but
the solution of PDE \ref{eq:chfk} is numerically cumbersome. 
A practical difficulty in
using the $\Omega$-expansion approach comes from the fact that
Eq.~\ref{eq:chfk} is a $(1+2)$ PDE, which does not seem to be a significant
simplification from the master equation, Eq. \ref{eq:m1}. In particular,
similar amount of numerical effort is needed to obtain solutions for
Eqs.~\ref{eq:m1} and \ref{eq:chfk}. This is part of the reason we did not try
to obtain the distribution from the $\Omega$-expansion. Therefore, we
used in the main text only the moments calculated from the $\Omega$-expansion.

\section*{Acknowledgments}

We thank Peter G. Wolynes for stimulating discussions. This work was supported
by R.J. Reynolds Excellence Junior Faculty Award.
 
\bibliography{../../biophys,../../nonlind}

\begin{thebibliography}{10}

\bibitem{st02gomp}
B.~D. Gomperts, I.~M. Kramer, and P.~E.~R. Tatham,
\newblock {\em Signal Transduction},
\newblock Academic Press, San Diego, 2002.

\bibitem{den95p}
D.~Bray,
\newblock Nature {\bf 376}, 307 (1995).

\bibitem{math02hei}
R.~Heinrich, B.~G. Neel, and T.~A. Rapoport,
\newblock Molecular Cell {\bf 9}, 957 (2002).

\bibitem{opt04cha}
M.~Chaves, E.~D. Sontag, and R.~J. Dinerstein,
\newblock J. Phys. Chem. B {\bf 108}, 15311 (2004).

\bibitem{han01ex}
D.~Hansel and G.Mato,
\newblock Phys. Rev. Lett. {\bf 86}, 4175 (2001).

\bibitem{nick04ph}
N.~I. Markevich, J.~B. Hoek, and B.~N. Kholodenko,
\newblock J. Cell Bio. {\bf 164}, 353 (2004).

\bibitem{h03ecoli}
K.~C. Huang, Y.~Meir, and N.~S. Wingreen,
\newblock Proc. Natl. Acad. Sci. USA {\bf 100}, 12724 (2003).

\bibitem{wein05st}
L.~S. Weinberger, J.~C. Burnett, J.~E. Toettcher, A.~P. Arkin, and D.~V.
  Schaffer,
\newblock Cell {\bf 122}, 169 (2005).

\bibitem{cao04eff}
Y.~Cao, H.~Li, and L.~Petzold,
\newblock J. Chem. Phys. {\bf 121}, 4059 (2004).

\bibitem{tan04mod}
T.~C. Meng, S.~Somani, and P.~Dhar,
\newblock In Silico Bio. {\bf 4}, 0024 (2004).

\bibitem{tur04st}
T.~E. Turner, S.~Schnell, and K.~Burrage,
\newblock Comput. Biol. Chem. {\bf 28}, 165 (2004).

\bibitem{dm05del}
D.~Brastsun, D.~Volfson, L.~S. Tsimring, and J.~Hasty,
\newblock Proc. Natl. Acad. Sci. USA {\bf 102}, 14593 (2005).

\bibitem{muk04st}
M.~Thattai and A.~van Oudenaarden,
\newblock Genetics {\bf 167}, 523 (2004).

\bibitem{jef02tr}
J.~Hasty and J.~J. Collins,
\newblock Natr. Genetics {\bf 31}, 13 (2002).

\bibitem{sasai03stch}
M.~Sasai and P.~G. Wolynes,
\newblock Proc. Natl. Acad. Sci. USA {\bf 100}, 2374 (2003).

\bibitem{muk02att}
M.~Thattai and A.~van Oudenaarden,
\newblock Biophys. J. {\bf 82}, 2943 (2002).

\bibitem{rao02n}
C.~V. Rao, D.~M. Wolf, and A.~P. Arkin,
\newblock Nature {\bf 420}, 231 (2002).

\bibitem{hol05st}
D.~Holcman and Z.~Schuss,
\newblock J. Chem. Phys. {\bf 122}, 114710 (2005).

\bibitem{coh05fl}
E.~Cohen, D.~A. Kesseler, and H.~levine,
\newblock Phys. Rev. Lett. {\bf 94}, 158302 (2005).

\bibitem{walcz04stgn}
A.~M. Walczak, M.~Sasai, and P.~G. Wolynes,
\newblock Biophys. J. {\bf 88}, 828 (2005).

\bibitem{paul00st}
J.~Paulsson, O.~G. Berg, and M.~Ehrenberg,
\newblock Proc. Natl. Acad. Sci. USA {\bf 97}, 7148 (2000).

\bibitem{paul00rd}
J.~Paulsson and M.~Ehrenberg,
\newblock Phys. Rev. Lett. {\bf 84}, 5447 (2000).

\bibitem{sh05noise}
T.~Shibata and K.~Fujimoto,
\newblock Proc. Natl. Acad. Sci. USA {\bf 102}, 331 (2005).

\bibitem{bark99cir}
N.~Barkai and S.~Leibler,
\newblock Nature {\bf 403}, 267 (1999).

\bibitem{kurt95st}
K.~Wiesenfeld and F.~Moss,
\newblock Nature {\bf 373}, 33 (1995).

\bibitem{jos01n}
J.~M.~G. Vilar, H.~Y. Kueh, N.~Barkai, and S.~Leibler,
\newblock Proc. Natl. Acad. Sci. USA {\bf 99}, 5988 (2002).

\bibitem{kor04var}
E.~Korobkova, T.~Emonet, J.~M.~G. Vilar, T.~S. Shimizu, and P.~Cluzel,
\newblock Nature {\bf 428}, 574 (2004).

\bibitem{wang94sur}
J.~Wang and P.~Wolynes,
\newblock Chem. Phys. {\bf 180}, 141 (1994).

\bibitem{wang96ins}
J.~Wang and P.~Wolynes,
\newblock J. Phys. Chem. {\bf 100}, 1129 (1996).

\bibitem{sm05inf}
V.~N. Smelyankiy, D.~G. Luchinsky, A.~Stefanovska, and P.~V.~E. McClintock,
\newblock Phys. Rev. Lett. {\bf 94}, 98101 (2005).

\bibitem{van92st}
N.~G. van Kampen,
\newblock {\em Stochastic processes in physics and chemistry},
\newblock North Holland Personal Library, Amsterdam, 1992.

\bibitem{gar02han}
C.~W. Gardiner,
\newblock {\em Handbook of stochastic methods},
\newblock Springer, New York, 2002.

\bibitem{ris84fok}
H.~Risken,
\newblock {\em The Fokker-Planck Equation},
\newblock Springer, Berlin, 1984.

\bibitem{wang02}
H.~Wang, C.~S. Peskin, and T.~C. Elston,
\newblock J. theor. Biol {\bf 221}, 491 (2003).

\bibitem{gill77ext}
D.~T. Gillespie,
\newblock J. Phys. Chem. {\bf 81}, 2340 (1977).

\bibitem{gil01app}
D.~T. Gillespie,
\newblock J. Chem. Phys. {\bf 115}, 1716 (2001).

\bibitem{jer05sim}
J.~S. van Zon and P.~R. ten Wolde,
\newblock Phys. Rev. Lett. {\bf 94}, 128103 (2005).

\bibitem{lin04hay}
F.~Hayot and C.~Jayaprakash,
\newblock Phys. Bio. {\bf 1}, 205 (2004).

\bibitem{tao05st}
Y.~Tao, Y.~Jia, and T.~G. Dewey,
\newblock J. Chem. Phys. {\bf 122}, 124108 (2005).

\bibitem{elf03fast}
J.~Elf and M.~Ehrenberg,
\newblock Genome Research {\bf 13}, 2475 (2003).

\bibitem{gil00lan}
D.~T. Gillespie,
\newblock J. Chem. Phys. {\bf 113}, 297 (2000).

\bibitem{gen04ch}
P.-G. de~Gennes,
\newblock Eur. Biophys. J. {\bf 33}, 691 (2004).

\bibitem{shnerb01aut}
N.~M. Shnerb, E.~Bettelheim, Y.~Louzoun, O.~Agam, and S.~Solomon,
\newblock Phys. Rev. E {\bf 63}, 021103 (2001).

\bibitem{gil03st}
M.~Rathinam, L.~R. Petzold, Y.~Cao, and D.~T. Gillespie,
\newblock J. Chem. Phys. {\bf 119}, 12784 (2003).

\bibitem{jac04b}
J.~Puchalka and A.~M. Kierzek,
\newblock Biophys. J. {\bf 86}, 1357 (2004).

\bibitem{gil03imp}
D.~T. Gillespie and L.~R. Petzold,
\newblock J. Chem. Phys. {\bf 119}, 8229 (2003).

\bibitem{eric02app}
E.~L. Haseline and J.~B. Rawlings,
\newblock J. Chem. Phys. {\bf 117}, 6959 (2002).

\bibitem{vas04ad}
K.~Vasudeva and U.~S. Bhalla,
\newblock Bioinformatics {\bf 20}, 78 (2004).

\bibitem{puc04bdg}
J.~Puchalka and A.~M. Kierzek,
\newblock Biophys. J. {\bf 86}, 1357 (2004).

\bibitem{keen05len}
J.~P. Keener,
\newblock J. Theor. Biol. {\bf 234}, 263 (2005).

\bibitem{ku01sel}
H.~Kuthan,
\newblock Prog. Biophys. Mol. Biol. {\bf 75}, 1 (2001).

\bibitem{keen01df}
J.~P. Keener,
\newblock Bull. Math. Biol. {\bf 63}, 625 (2001).

\bibitem{lem05st}
C.~Lemerle, B.~D. Ventura, and L.~Serrano,
\newblock FEBS Lett. {\bf 579}, 1789 (2005).

\bibitem{kul04pat}
R.~V. Kulkarni, K.~C. Huang, M.~Kloster, and N.~S. Wingreen,
\newblock Phys. Rev. Lett. {\bf 93}, 228103 (2004).

\bibitem{th01in}
M.~Thattai and A.~van Oudenaarden,
\newblock Proc. Natl. Acad. Sci. {\bf 98}, 8614 (2001).

\bibitem{kie01eff}
A.~M. Kierzek, J.~Zaim, and P.~Zielenkiewicz,
\newblock J. Biol. Chem. {\bf 276}, 8165 (2001).

\bibitem{sw02in}
P.~S. Swain, M.~B. Elowitz, and E.~D. Siggia,
\newblock Proc. Natl. Acad. Sci. {\bf 99}, 12795 (2002).

\bibitem{oz02reg}
E.~M. Ozbudak, M.~Thattai, I.~Kurtser, A.~D. Grossman, and A.~van Oudenaarden,
\newblock Nature Genet. {\bf 31}, 69 (2002).

\bibitem{pugh92r}
J.~E.~N.~Pugh and T.~D. Lamb,
\newblock Biochim. Biophys. Acta {\bf 1141}, 111 (1993).

\bibitem{sch02com}
B.~Schoeberl, C.~Eichler-Jonsson, E.~D. Gilles, and G.~M{\"u}ler,
\newblock Nat. Biotechnol. {\bf 20}, 370 (2002).

\bibitem{bur01act}
M.~E. Burns and D.~A. Baylor,
\newblock Annu. Rev. Neurosci. {\bf 24}, 779 (2001).

\bibitem{sch95pht}
D.~M. Schneeweis and J.~L. Schnapf,
\newblock Science {\bf 268}, 1053 (1995).

\bibitem{det00en}
P.~B. Detwiler, S.~Ramanathan, A.~Sengupta, and B.~I. Shraiman,
\newblock Biophys. J. {\bf 79}, 2801 (2000).

\bibitem{rk98or}
F.~Rieke and D.~A. Baylor,
\newblock Biophys. J. {\bf 75}, 1836 (1998).

\bibitem{thr04dif}
R.~G. Thorne and S.~Hrab{\v{e}}tov{\'a},
\newblock J. Neurophysiol. {\bf 92}, 3471 (2004).

\bibitem{el99prt}
M.~B. Elowitz, M.~Surette, P.~Wolf, J.~Stock, and S.~Leibler,
\newblock J. Bacteriol. {\bf 181}, 197 (1999).

\bibitem{ince}
E.~L. Ince,
\newblock {\em Ordinary Differential Equations},
\newblock Dover, New York, 1956.

\bibitem{bla03noi}
W.~J. Blake, M.~Kaern, C.~R. Cantor, and J.~J. Collins,
\newblock Nature {\bf 422}, 633 (2003).

\bibitem{th01st}
T.~B. Kepler and T.~C. Elston,
\newblock Biophys. J. {\bf 81}, 3116 (2001).

\bibitem{jas04fl}
J.~R. Pirone and T.~C. Elston,
\newblock J. Theor. Biol. {\bf 226}, 111 (2004).

\bibitem{bun03fl}
R.~Bundschuh, F.~Hayot, and C.~Jayaprakash,
\newblock Biophys. J. {\bf 84}, 1606 (2003).

\bibitem{hay04lin}
F.~Hayot and C.~Jayaprakash,
\newblock Phys. Biol. {\bf 1}, 205 (210).

\bibitem{nic77self}
G.~Nicolis and I.~Prigogine,
\newblock {\em Self-organization in nonequilibrium systems},
\newblock A Wiley-Interscience Publication, New York, 1977.

\bibitem{rei98st}
L.~E. Reichl,
\newblock {\em A Modern Course in Statistical Physics},
\newblock John Wiley, New York, 1998.

\bibitem{bar98bac}
N.~Barkai and S.~Leibler,
\newblock Nature {\bf 393}, 18 (1998).

\bibitem{eyink96act}
G.~L. Eyink,
\newblock Phys. Rev. E {\bf 54}, 3419 (1996).

\end{thebibliography}

\end{document}